\documentclass[preprint]{elsarticle}

\usepackage{hyperref}
\hypersetup{colorlinks=true, breaklinks=true}
\usepackage{color,soul}
\usepackage{multirow}
\usepackage{nicefrac}
\usepackage{amsmath}
\usepackage{amssymb}
\usepackage{microtype}
\usepackage{lipsum}

\journal{Nucl.\ Instrum.\ Methods Phys.\ Res., Sect.\ A}

\newcommand{\nuc}[2]{${}^{#1}\mathrm{#2}$}

\newcommand{\rxn}[6]{${}^{#1}\mathrm{#2}(#3,#4){}^{#5}\mathrm{#6}$}
\newcommand{\Nean}{${}^{22}\mathrm{Ne}(\alpha,n){}^{25}\mathrm{Mg}$}

\newcommand{\edit}[1]{{#1}}
\newcommand{\editbf}[1]{{#1}}

\begin{document}

\begin{frontmatter}

\title{{DEMAND}: {A}n organic glass scintillator array for direct measurements of $(\alpha,n)$ and $(p,n)$ reactions in inverse kinematics}

\author[SMU]{G.~Christian\corref{mycorrespondingauthor}}
\ead{Greg.Christian@smu.ca}
\cortext[mycorrespondingauthor]{Corresponding author}
\author[SMU,Surrey,TRIUMF]{B.~J.~Reed}
\author[Surrey]{G.~Lotay}
\author[Surrey]{G.~Bartram}
\author[SMU]{S.~Blackmore}
\author[Surrey]{D.~T.~Doherty}
\author[TRIUMF]{A.~Katrusiak\fnref{AlexPresent}}
\fntext[AlexPresent]{Present address: Department of Physics \& Astronomy, University of Victoria, Victoria, BC V8W~3P6, Canada}
\author[Surrey]{J.~Henderson}
\author[TRIUMF]{D.~A.~Hutcheon}
\author[TRIUMF]{A.~Lennarz}
\author[TRIUMF,UVic]{M.~Loria}
\author[Surrey]{J.~O'Neill\fnref{JoeyPresent}}
\fntext[JoeyPresent]{Present address: School of Physics \& Astronomy, University of Birmingham, Birmingham B15~2TT, UK}
\author[Surrey]{C.~O'Shea}
\author[Surrey]{C.~J.~Paxman}
\author[TRIUMF,UVic]{C.~Ruiz}
\author[TRIUMF]{L.~Wagner\fnref{louispresent}}
\fntext[louispresent]{Present address: Facility for Rare Isotope Beams, Michigan State University, East Lansing, MI 48824 USA}
\author[Surrey]{M.~Williams}

\address[SMU]{Department of Astronomy \& Physics, Saint Mary's University, Halifax, NS B3H~3C3, Canada}
\address[Surrey]{School of Mathematics \& Physics, University of Surrey, Guildford GU2~7XH, UK}
\address[TRIUMF]{TRIUMF, 4004 Wesbrook Mall, Vancouver, BC B6T~2A3, Canada}
\address[UVic]{Department of Physics \& Astronomy, University of Victoria, Victoria, BC V8W~3P6, Canada}

\begin{abstract}
	\edit{We have developed a new array of organic glass scintillator detectors, consisting of up to {20} $3~\mathrm{cm}\times 3~\mathrm{cm}\times 3~\mathrm{cm}$ scintillator cubes, each coupled to a photomultiplier tube. The array is sensitive to neutrons with energies ranging from a few hundred keV up to several tens of MeV. It was designed to be coupled with a recoil separator for inverse-kinematics direct measurements of astrophysical $(\alpha,n)$ or $(p,n)$ reactions, where it provides a prompt start signal for identifying recoil-neutron coincidences from the recoil time of flight. In this article, we present the results of both source and in-beam measurements characterizing the properties of a sub-set of eight detectors used for a measurement of the \Nean{} reaction in inverse kinematics. We report specifically on the detector's energy and timing resolution and pulse shape discrimination capabilities. From the widths of $\gamma$-ray Compton-edge peaks at 311, 341, 477, and 1062 keVee, we find that the energy resolution as a function of energy is well described by $	\mathrm{FWHM}(E)/E = \sqrt{A^2/E + B^2/E^2 + C^2},$ with $A=0.00(7)$, $B=0.0618(54)$, and $C=0.135(5)$.  From analyzing the time difference spectrum between nearest-neighbor $\gamma$-$\gamma$ coincidence events in the energy range ${\sim}250$-600~keVee, produced by a \nuc{22}{Na} source, we find a FWHM coincidence timing resolution of $910(40)$~ps, implying a single-detector time resolution of $640(30)$~ps. Using mixed neutron-$\gamma$-ray fields produced by both the in-beam \Nean{} reaction and a \nuc{241}{AmBe} source, we find that the pulse shape discrimination provides $n$-$\gamma$ separation with figure of merit $>1$ down to a threshold of around $100$~keVee. Additionally, we demonstrate that the total neutron yield below $100$~keVee can be extracted from two-Gaussian fits to energy-gated pulse-shape discrimination spectra.
	 We also discuss the results of Monte Carlo simulations to determine the array's neutron detection efficiency in its present configuration. Simulating the present \Nean{} measurements, we find a neutron singles detection efficiency of $1.56(45)\%$, with the uncertainty dominated by systematic effects related to the simulation's input parameters.}
\end{abstract}


\end{frontmatter}


\section{Introduction and Background}
\label{sec:Intro}

Neutron-emitting reactions such as $(\alpha,n)$ or $(p,n)$ play an important role in a variety of astrophysical environments, with impacts on the origin of the elements. 
%
{{In order to better constrain nucleosynthesis predictions, measurements of key astrophysical reaction rates are required, which motivates the construction of new neutron detector arrays.}}
%
\edit{One {{astrophysical}} example where {{neutron-emitting reactions are important}} is the origin of the elements heavier than iron, which cannot be produced through conventional stellar fusion due to the decreasing binding energy per nucleon and the increasing Coulomb barrier. Instead, the vast majority of the heavy elements are made through neutron capture processes, specifically either the ``rapid'' ($r$) or ``slow'' ($s$) neutron capture processes. The $r$ process occurs in astrophysical sites with high neutron flux, such as neutron star mergers, with rapid sequences of $(n,\gamma)$ reactions reaching far from stability before $\beta^-$ decaying to produce heavy stable nuclides. Some astronomical observations, e.g.\ of metal-poor halo stars, show enhanced abundances of the lighter $r$-process elements (Sr--Ag) relative to solar $r$-process abundance patterns \cite{Bliss2018, Frebel2018}. This indicates that there may be other nucleosynthesis processes producing these so-called ``weak $r$-process'' elements. One possible production site is the so-called ``$\alpha$ process,'' occurring in the $\nu$-driven winds following core-collapse supernovae.  In this scenario, $(n,\gamma)$ and $(\gamma,n)$ reactions are in equilibrium, and, due to the short timescale (tens of milliseconds), $\beta^-$ decays are too slow to compete with $(\alpha,n)$ reactions in driving the pathway of nucleosynthesis towards heavier elements. As a result, certain $(\alpha, n)$ reaction rates have a substantial impact on $\alpha$-process nucleosynthesis, with sensitivity studies demonstrating that $(\alpha,n)$ reactions in the Kr--Sr region, mostly involving short-lived radioactive nuclides, have the most significant impact \cite{Bliss2020}.}


\edit{Turning to the $s$ process, $(\alpha,n)$ reactions are important as the source of the neutron flux that is needed to drive the heavy-element synthesis. The $s$ process occurs in massive and asymptotic giant branch (AGB) stars, where elements heavier than Fe are produced through sequences of $(n,\gamma)$ reactions and $\beta^-$ decays, with a timescale between capture reactions on the order of years. Due to the long time between successive reactions, the $s$ process proceeds close to stability, thus producing a different abundance pattern than the $r$ process. As mentioned, $(\alpha,n)$ reactions, in particular \rxn{13}{C}{\alpha}{n}{16}{O} and \Nean{} provide the required neutron flux for the $s$ process to take place, and hence, the rates of these reactions have a substantial impact on $s$-process nucleosynthesis. Most notably, there are significant uncertainties on the \Nean{} reaction rate at $s$-process temperatures, stemming from disagreement between recent experiments on the strength of the key $E_{cm} = 704$~keV resonance strength \cite{Lugaro2023}.}

As a final example, both $(p,n)$ and $(\alpha,n)$ reactions, often on radioactive nuclides, play an important role in the production of long-lived $\gamma$-ray emitters during core-collapse supernovae, many of which are potentially observable using the next generation of $\gamma$-ray telescopes. \edit{For example, the \rxn{42,43}{K}{p}{n}{42,43}{Ca} reactions both have a strong impact on the production of potentially-observable \nuc{43}{K}. According to the sensitivity study by Hermansen \emph{et al.}, varying these reactions within existing uncertainties results in predicted \nuc{43}{K} abundance changes by factors 2.51 and 2.13 for the \nuc{43}{K} and \nuc{42}{K} reactions, respectively \cite{Hermansen2020}.}


\edit{Many of the most important $(\alpha,n)$ and $(p,n)$ reactions of interest for  astrophysics involve short-lived radioactive nuclides.} 
Likewise, key stable-beam reactions such as \Nean{} are often challenging to measure in the astrophysical energy regime. Inverse-kinematics experiments, \edit{where a heavy beam impinges on a light target, with the heavy reaction products detected in a recoil separator, offer a potential solution to both of these issues. This technique can be combined with radioactive beams to study reactions involving short-lived reactants that would otherwise not be possible. For stable-beam reactions, inverse kinematics offers a high degree of background suppression and different, often well characterized, systematic uncertainties as compared to normal-kinematics experiments.} 
In recent years, significant progress has been made measuring $(\alpha,n)$ and $(p,n)$ reactions in inverse kinematics. For example, multi-sampling ionization chambers have been used to detect the heavy-ion recoil products of $(\alpha,n)$ reactions \cite{Avila2017, Jayatissa2023, Blankstein2024}, and recoil separators have been used to detect the heavy-ion recoils in coincidence with secondary $\gamma$-rays \cite{Williams2025}.   A complementary technique currently under development is the coupling of a recoil separator with neutron detectors, which has previously been used to study $(p,n)$ reactions \cite{Gastis2021,Tsintari2025}. This method allows for the coincident detection of the heavy-recoil reaction products together with the neutrons. \edit{Unlike the technique of tagging secondary $\gamma$~rays, the recoil-neutron coincidence method can be used when neutron decay directly to the ground state is the dominant (or exclusive) de-excitation mechanism. For reactions where feeding to excited states dominates, the recoil-neutron method is less sensitive than the recoil-$\gamma$ method to the $\gamma$-ray decay branching ratios of the daughter, which are often unknown. In recoil-neutron measurements, these branching ratios only appear as a higher-order correction related to the neutron detection efficiency, whereas in recoil-$\gamma$ measurements the feeding probability to the tagged $\gamma$-ray lines is directly proportional to the extracted cross section.}

When measuring $(\alpha,n)$ or $(p,n)$ reactions using a recoil separator coupled to neutron detectors, a prompt timing signal is required from the neutrons in order to use time of flight to correlate neutron events with recoils detected in the focal plane of the separator. The typical neutron energies encountered in such measurements range from a few hundred keV up to approximately $20$~MeV, making organic scintillator detectors well suited for this application. These detectors measure scintillation light produced by $n$-$p$ scattering within the detector volume (as opposed to relying on neutron thermalization), thus providing the required prompt timing signal. While conventional plastic scintillators can be used, e.g.\ as in Ref.~\cite{Gastis2021}, detectors with $n$/$\gamma$ pulse shape discrimination (PSD) provide additional background suppression through the rejection of both genuine and random recoil-$\gamma$ coincidences. This is especially true when the neutron detectors are placed close to the reaction target, making neutrons and $\gamma$~rays indistinguishable based on time of flight.

In recent years, a number of new PSD-capable solid organic scintillators have been developed. These offer excellent PSD resolution, comparable with traditional liquid scintillators but without the liquid's inherent safety concerns. \edit{The alleviation of safety concerns is an advantage for detectors that travel between laboratories, and more generally to reduce administrative and regulatory burden.} Solid detectors are also well suited to compact layouts, which are often necessary due to extensive mechanical infrastructure surrounding the targets of recoil separators. One promising material is organic glass scintillator (OGS), which has excellent PSD, fast timing, and high light output \cite{Carlson2016, Carlson2017, Shin2019, Laplace2020, Ghia2021, Warburton2021, Pozzi2026}. \edit{These properties lead to significantly lower PSD thresholds than PSD-capable plastics such as EJ-276 \cite{Warburton2021}, which increases detection efficiency for neutrons below ${\sim}2$ MeV. This is an advantage in measurements such as \Nean{}, which requires detection of neutrons ranging from 0.7--1.8 MeV in energy, depending on the $(\alpha,n)$ resonance being studied.} 
Due to its glass formulation, detectors are also able to be fabricated into a variety of shapes, which makes it well suited for arrays mounted near the targets of recoil separators. \edit{Although only cube-shaped detectors are used in the present work, this feature could prove useful in future efforts.}

Motivated by measurements of $(\alpha,n)$ and $(p,n)$ reactions for nuclear astrophysics, we have developed a new array of organic glass scintillators targeted at inverse-kinematics measurements coupled with a recoil separator: the array for Direct Experimental Measurements of Astrophysical reactions using Neutron Detectors (DEMAND). 
DEMAND consists of up to {20}, $3~\mathrm{cm}\times 3~\mathrm{cm}\times 3~\mathrm{cm}$ OGS cubes, each coupled to a Hammamatsu H6533 photomultiplier tube (PMT). \edit{The cubes were provided by Blueshift Optics, LLC  \cite{BlueshiftOptics}.} \edit{DEMAND} is designed to be coupled with the DRAGON recoil separator in the ISAC-I facility at TRIUMF \cite{Hutcheon2003}, \edit{where it partially or fully replaces} 
the Bismuth Germanate (BGO) $\gamma$-ray detectors surrounding the reaction target. \edit{This allows} recoil-$n$ coincidence measurements to be made using \edit{DRAGON, expanding} its measurement program to include $(\alpha,n)$ and $(p,n)$ reactions. 
\edit{Example DEMAND experiments, which we intend to complete in the next few years, include measuring the  strength of the key, $E_{cm}= 704$~keV, \Nean{} resonance, as well as the \rxn{38}{Ar}{\alpha}{n}{41}{Ca} and \rxn{43}{K}{p}{n}{43}{Ca} cross sections within the Gamow window for core-collapse supernova nucleosynthesis.} 
\edit{The neutron-detection requirements for the planned measurement program are heavily influenced by reaction kinematics, and hence the array, or a sub-set thereof, is intended to be deployed in a customized configuration for each experiment.} In the present article, we report on the performance of a \edit{sub-set of eight detectors} in both source tests and a commissioning experiment measuring the $E_{cm} = 1213$~keV resonance in the \Nean{} reaction. The present article focuses only on the performance of the neutron detectors, discussing their energy, timing, and PSD performance, as well as simulated detection efficiencies. \edit{The article also demonstrates a method for extracting total neutron singles counts from the data.} \edit{This represents the first phase of detector characterization work, and future characterization efforts, centered around measurements at a neutron-beam facility, will be reported in a later technical article.} \edit{Similarly, the} complete results of the commissioning measurement, including a recoil-$n$ coincidence analysis, will be reported in a future article.

\section{Experiment}
\label{sec:expt}

The experiment was performed at the ISAC-I facility at TRIUMF \cite{Laxdal2003400}. A \nuc{22}{Ne}$^{(4+)}$ beam was extracted from the Pantechnik Supernanogan ECR source \cite{jayamanna:02C711} and delivered to the DRAGON facility with an intensity of ${\sim}1$~e$\mu$A and an energy of $E/A = 362.2(5)$~keV. The DRAGON recirculating gas target was filled with He gas at an average pressure of $6.5(1)$~mbar at a temperature of 303.2(1) K. These beam and target conditions approximately centered the $E_{cm} = 1213$~keV resonance in the gas target. For the present set of runs, DRAGON was tuned to accept \nuc{25}{Mg}$^{(4+)}$ recoils resulting from the \Nean{} reaction. Since the maximum \nuc{25}{Mg} cone angle is larger than DRAGON's ${\sim}20$ mrad angular acceptance, as well as the ${\sim}4\%$ energy acceptance, the separator was tuned to select the \nuc{25}{Mg} emitted in coincidence with forward-going neutrons in the laboratory frame.

\begin{figure}
\centering
\centerline{\includegraphics[width=90mm]{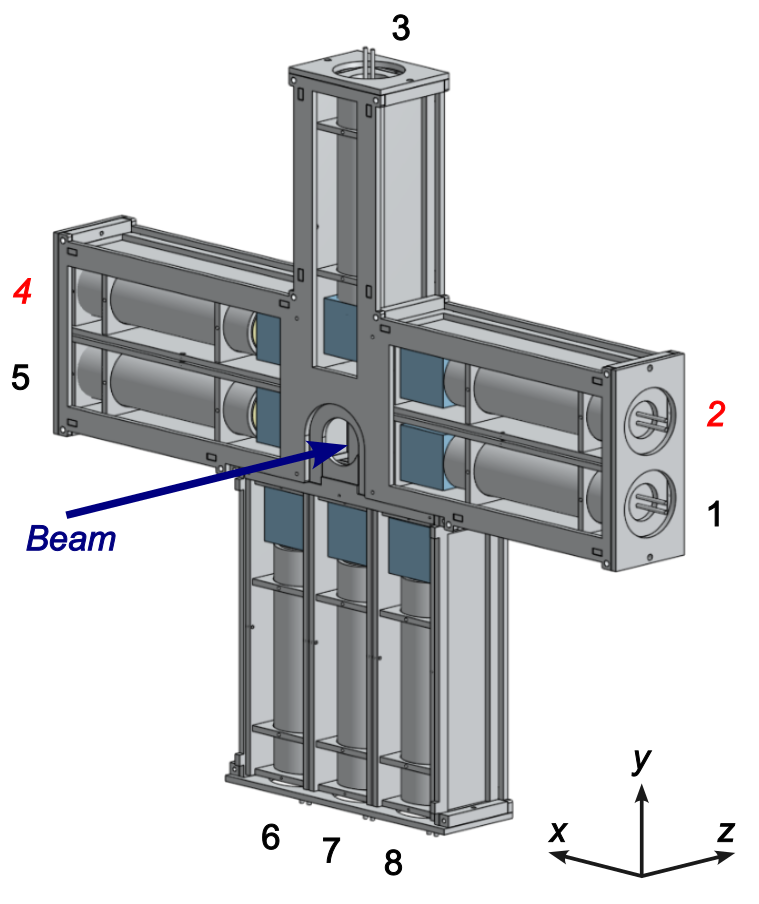}}
\caption{
	CAD rendering of the DEMAND detector assembly. \edit{The upstream faces of the detector housing have been made transparent to show the OGS cubes (light blue color), PMTs, and collars inside.} Each detector is labeled with a detector number corresponding to those listed in Table~\ref{tab:DetectorPositions}. \edit{Detectors 2 and 4 are labeled with red italic text to indicate that they were inactive during the initial long $(\alpha,n)$ runs, as explained in the text.}
}
\label{fig:DEMAND}
\end{figure}

The detection system surrounding the DRAGON target was modified to detect the forward-going neutrons from ${}^{22}\mathrm{Ne}(\alpha, n){}^{25}\mathrm{Mg}$. Ordinarily, a compact array of 30 BGO detectors surrounds the gas target to detect the $\gamma$-rays resulting from radiative capture reactions. In the present experiment, the outermost forward-angle detectors were removed to avoid shadowing neutrons emitted from the target, while the remainder of the BGO array remained in place. For neutron detection, an array of 8 DEMAND detectors was placed in a ring-like geometry, with the center of each detector located $11.66$~cm downstream of the center of the gas target. The placement of the array was chosen to maximize the $n$-\nuc{25}{Mg} coincidence efficiency, while taking into account space constraints along the DRAGON beamline.  Each scintillator cube was wrapped in six layers of 0.075~mm thick white PTFE tape, a diffuse reflector, along with one layer of black electrical tape, for light tightness. Each wrapped scintillator, together with its PMT, was situated in a thin-walled (3.175 mm) aluminum casing. A custom-designed aluminum mounting frame secured the assembly to the beamline.
Figure~\ref{fig:DEMAND} shows a CAD drawing of the DEMAND array, including the detector casings and mounting frame, and Table~\ref{tab:DetectorPositions} lists the horizontal and vertical positions of each detector, relative to the center of the gas target.  During the initial long set of $(\alpha, n)$ runs, whose analysis is reported in the present article, two detectors were not fully operational (numbers 2 and 4 in Table~\ref{tab:DetectorPositions}). These were fixed later in the beam time, before a number of shorter runs were taken with differing DRAGON field settings. The present analysis is limited to the sub-set of six detectors that were operational during the initial runs.

\begin{table}
\begin{center}
\caption{Positions of the center of each OGS detector in the DEMAND array. The coordinate system used to label the positions is shown in Figure~\ref{fig:DEMAND}. Positions are given relative to the center of the gas target, and $y=0$ is the beam height. The $z$ position of each detector is 11.66~cm.
}
\label{tab:DetectorPositions}
\begin{tabular}{ccc}
\hline
Number & $x$ (cm) & $y$ (cm)\\
\hline
 1 & $ -5.27 $  & $0$      \\
 2 & $ -5.27 $  & $+4.45$  \\
 3 & $ 0     $  & $+5.27$  \\
 4 & $ +5.27 $  & $+4.45$  \\
 5 & $ +5.27 $  & $0$      \\
 6 & $ +4.45 $  & $-5.27$  \\
 7 & $ 0     $  & $-5.27$  \\
 8 & $ -4.45 $  & $-5.27$  \\
\hline
\end{tabular}
\end{center}
\end{table}

\begin{figure*}
\centering
\centerline{\includegraphics[width=140mm]{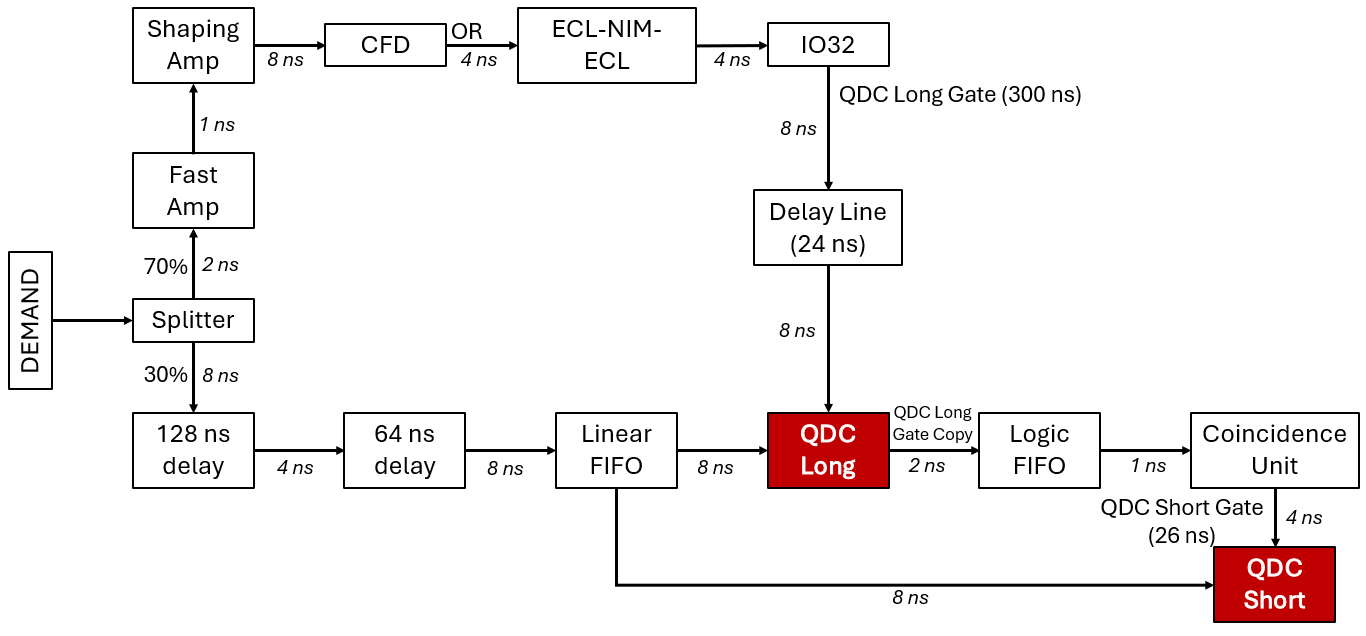}}
\caption{
	Schematic of the DAQ triggering and gating scheme for the DEMAND detectors.
}
\label{fig:DAQ}
\end{figure*}

The DEMAND detectors were read out using a modified version of the standard DRAGON BGO electronics \edit{\cite{Christian2014}}. The PMT anode signal was first sent to a $70/30$ passive splitter, with the $70\%$ branch send to a CAEN V812 constant fraction discriminator (CFD). The individual CFD outputs were sent to a V1190 time-to-digital converter (TDC), while the \edit{logical \textsc{or}} output was used to generate the data acquisition trigger. The $30\%$ branch was first sent through a cable delay to synchronize it with the arrival time of the system trigger. After the delay, the signal was sent to a LeCroy model 428 linear fan-in/fan-out (FIFO) to further split the signal. Each of the FIFO outputs was sent to a CAEN V792 charge-to-digital converter (QDC), which integrated the pulse over a carefully selected time gate.  The gates were produced from copies of the system trigger, with the gate lengths and delays tuned to overlap the desired portions of the anode signal. The first QDC received a gate width of $300$~ns (``long gate''), integrating the entire anode pulse, along with a small amount of baseline prior to the leading edge. The second pulse received a gate width of ${\sim}25$~ns (``short gate''), with the majority of the pulse tail excluded from the gate window. As described in Section~\ref{sec:PSD}, charge comparison between the long and short gates was used to perform PSD between incident neutron and $\gamma$-ray events. Figure~\ref{fig:DAQ} shows a schematic of the DAQ trigger logic and gating scheme.

The PMT high voltages were set to ${\sim}2$ kV, and for the present experiment the detectors were gain matched (and later calibrated offline) using the 477~keV Compton edge from a \nuc{137}{Cs} source.  Data were also taken with a ${}^{241}$AmBe source, which provided a mixed $n$-$\gamma$ field to evaluate the  PSD capabilities of the detectors. In a later iteration of the setup (with slightly different detector gains), source runs were taken using \nuc{7}{Be}, \nuc{22}{Na} and \nuc{137}{Cs} sources (the \nuc{7}{Be} had been deposited along the beamline in an unrelated experiment). Together, these provided incident $\gamma$-rays at 478, 511, 662, and 1275~keV, including coincident 511~keV $\gamma$-$\gamma$ pairs from \nuc{22}{Na} positron annihilation. The data from these source measurements were used to characterize both the energy and timing resolution of the detectors, as discussed in Sections~\ref{sec:EnergyRes} and \ref{sec:TimeRes}.

\section{Data Analysis}
\label{sec:Analysis}

\subsection{Energy Resolution}
\label{sec:EnergyRes}

For organic scintillator neutron detectors, the deposited energy depends on the $n$-$p$ scattering angle, which is isotropic in the center of mass frame. As a result, limited spectroscopic information can be obtained from energy spectra \editbf{\cite{Knoll2010}}. In a typical direct $(\alpha,n)$ or $(p,n)$ measurement, the energy signal is used minimally, for example only to set an offline threshold for background rejection. However, understanding the energy response of the neutron detectors is important for modeling their detection efficiency, in particular for understanding the number of events removed by detector thresholds. For this reason, we have characterized the resolution of the DEMAND detectors as a function of deposited energy using $\gamma$-rays of varying energies.

\begin{figure}
\centering
\centerline{\includegraphics[width=90mm]{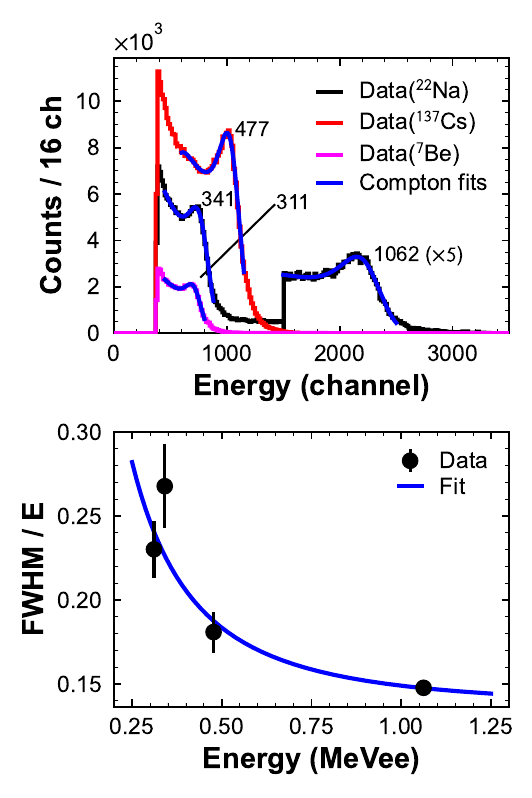}}
\caption{
	Top panel: Sample Compton edge energy spectra for a single DEMAND detector. Measured data from \nuc{7}{Be}, \nuc{137}{Cs}, and \nuc{22}{Na} sources, along with least-squares fits to Equation~\ref{eq:CE}, are shown as indicated in the legend. The various peaks in the three spectra are labeled with the electron recoil energy from $180^{\circ}$ Compton scattering.
	Bottom panel: Average FWHM energy resolution vs.\ deposited energy. The black filled circles mark the average FWHM/$E$ across all detectors, and the error bars denote the standard deviation of the same. The blue curve represents the results of the fit to Equation~\ref{eq:FWHMvsE}.
}
\label{fig:compton_fit}
\end{figure}

The energy resolution measurements were performed using Compton-scattering spectra from both \nuc{22}{Na} and \nuc{137}{Cs} calibration sources, as well as \nuc{7}{Be} that had been deposited near the target in an unrelated experiment. Together, these provide incident $\gamma$~rays with energies of 478, 511, 662, and 1275~keV, which have respective Compton edges at 311, 341, 477, and 1062~keVee \editbf{\cite{Tilley2002,Basunia2015,Browne2007}}. The measured energy spectra from each of these sources are shown in the top panel of Figure~\ref{fig:compton_fit}. Due to detector resolution, each of the Compton edges appears as a Gaussian-like peak on top of the Compton-continuum background.  To extract the Compton edge positions, the peaks were fit with a response function, $R(E)$, given by the convolution of an idealized response, $r(E)$, with a Gaussian:
\begin{equation}\label{eq:CE}
	R(E) = 
	\frac{1}{\sqrt{2\pi}\sigma} \int_{-\infty}^{\infty}{
		r(E^\prime)\exp\left[
			-\frac{(E-E^\prime)^2}{2\sigma^2}
		\right] \mathrm{d}E^\prime
	}.
\end{equation}
For the idealized response, $r(E)$, we approximated the expected Klein-Nishina shape below the Compton edge as a second order polynomial divided by a square-root function:
\begin{equation}
r(E) = \begin{cases}
			\dfrac{a + b(E_c-E) + c(E_c-E)^2}{
				\sqrt{E_c-E}}, & E\leq E_c\\
			0,              & E > E_c.
		\end{cases}
\end{equation}
Here, $E_c$ is the location of the Compton edge, and $a$, $b$, and $c$ are free parameters describing the shape of the polynomial. The convolution integral in Equation~\ref{eq:CE} was solved numerically. The response function $R(E)$ provides a good match to the measured energy-deposition spectra in a local window surrounding each of the four Compton edges, with the best-fit parameters robust to variations in the window size and starting parameters. \editbf{Similar techniques to calibrate organic scintillators using Compton edges have been used by other groups, e.g.\ Refs.\ \cite{Kudomi1999,Stevanato2011}.}

Figure~\ref{fig:compton_fit} (top panel) compares the measured energy spectra from a single detector with the results of a least-squares fit to Equation~\ref{eq:CE}. The $\sigma$ parameter extracted from each of the fits provides a measure of the detector resolution at the corresponding Compton-edge energy, while the $E_c$ parameter is used to calibrate the energy scale.  To characterize the average energy resolution of the detectors, we calculated the mean and standard deviation of similar fits across all detectors and plotted the average resolution vs.\ the Compton-edge energy. This is shown in Figure~\ref{fig:compton_fit} (bottom panel), where the resolution is expressed as the full-width at half maximum ($\mathrm{FWHM} = 2.355\sigma$) divided by the deposited energy. \edit{The error bars in the plot represent the standard deviation of the resolutions across all detectors in the sample.} The energy dependence of $\mathrm{FWHM}/E$ is reasonably well described by the form given by Knoll \cite{Knoll2010},
\begin{equation}\label{eq:FWHMvsE}
	\mathrm{FWHM}(E)/E = 
\sqrt{A^2/E + B^2/E^2 + C^2},
\end{equation}
with best-fit parameters \edit{$A=0.00(7)$, $B=0.0618(54)$, and $C=0.135(5)$}. 
\edit{We emphasize that these results require extrapolation down to threshold energies, which could impact the simulated detection efficiency discussed in Section~\ref{sec:MC}. In the future, we plan to extend our energy resolution measurements to include sources such as \nuc{57}{Co}, which emits a $122$~keV $\gamma$~ray.} 



\begin{figure}
	\centering
	\centerline{\includegraphics[width=90mm]{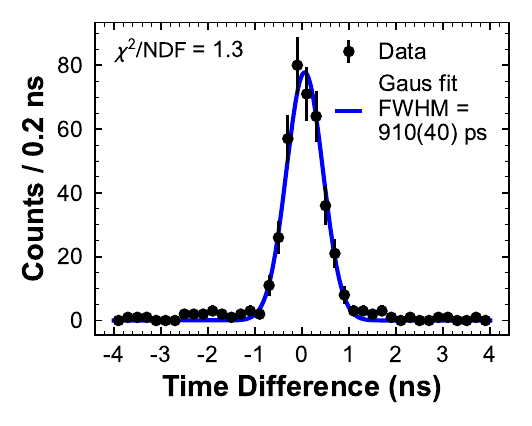}}
	\caption{
		\edit{Time difference spectrum for nearest-neighbor $\gamma$-$\gamma$ coincidence events produced by the \nuc{22}{Na} calibration source. The black filled circles represent the measured data, and the solid blue curve represents a least-squares fit to a Gaussian function. The Gaussian FWHM and $\chi2/\mathrm{NDF}$ are given in the legend.}
	}
	\label{fig:tdiff}
\end{figure}

\subsection{Timing Resolution}
\label{sec:TimeRes}
Good timing resolution is an important characteristic of neutron detectors used in inverse-kinematics direct measurements. Typically, the primary recoil-neutron coincidence signal is the recoil time of flight through the entire length of the separator, with the start signal provided by the neutron detectors. \editbf{This is similar to the recoil-$\gamma$ time of flight coincidence signal used in radiative capture measurements, e.g.\ Refs.~\cite{Bishop2003,Ruiz2006,Lotay2016}.} The coincidence timing spread is often dominated by the spread in recoil velocities, which can be as small as a few ns in some experiments. For example, for the present $E_{cm}=1.213$~MeV \Nean{} resonance, the recoil time-of-flight spread is around $10$ ns FWHM. As a result, neutron-detector time resolutions on the order of a ns or better are required to maximize the coincidence detection (and background rejection) capabilities of the separator. Good timing resolution may also prove valuable for other applications. For example, if the neutron flight path is long enough and the reaction occurs in a localized region of the gas target, the time of flight between the neutrons and the arrival of the beam bunch (measured from the accelerator RF signal) may be used to infer neutron energies and hence separate excited-state channels populated in the reaction. Similarly, neutron-RF timing could be used to determine narrow-resonance energies, as has recently been done at DRAGON for radiative capture reactions \cite{Christian2025}.

\edit{The timing resolution of the DEMAND detectors was evaluated using multiple-scattered $\gamma$~ray events generated with the \nuc{22}{Na} source. The source had an activity of 69~kBq and was placed on the upstream face of the detector assembly while it was installed on the beamline, just above the beampipe on the vertical axis. Referencing Figure~\ref{fig:DEMAND}, the source would be at approximately $x=0$ and centered between detectors $1$ and $2$ in $y$. For the timing analysis, we selected only nearest-neighbor coincidences with the lower-energy hit $>250$~keVee and the difference in energy between the coincident hits $<50$~keVee.  The resulting time-difference spectrum is shown in Figure~\ref{fig:tdiff}. The data in the figure are corrected for timing-offset differences between the various detectors, and they are shifted so that the mean appears at zero. We performed a least-squares fit with a Gaussian function, with the resulting $\mathrm{FWHM} = 910(40)$~ps ($\chi^2/\mathrm{NDF} = 1.3$). The FWHM from the fit represents the average coincidence timing resolution. Assuming equal contributions from each detector, then the FWHM timing resolution for a single detector would be 640(30)~ps. This is sufficient for coincidence tagging in the proposed $(\alpha,n)$ and $(p,n)$ direct measurements for which the array was designed. The energy range spanned by our present measurements, approximately 250-600~keVee with the analysis cuts, is near the lower end of the expected light output for neutrons produced during our proposed $(\alpha,n)$ and $(p,n)$ measurements. In the future, we plan additional coincidence timing measurements spanning a wider range of deposited energies, which will allow us to investigate if there is any significant dependence of the timing resolution on the deposited energy.}

\subsection{Pulse Shape Discrimination}
\label{sec:PSD}


\begin{figure*}
\centering
\centerline{\includegraphics[width=140mm]{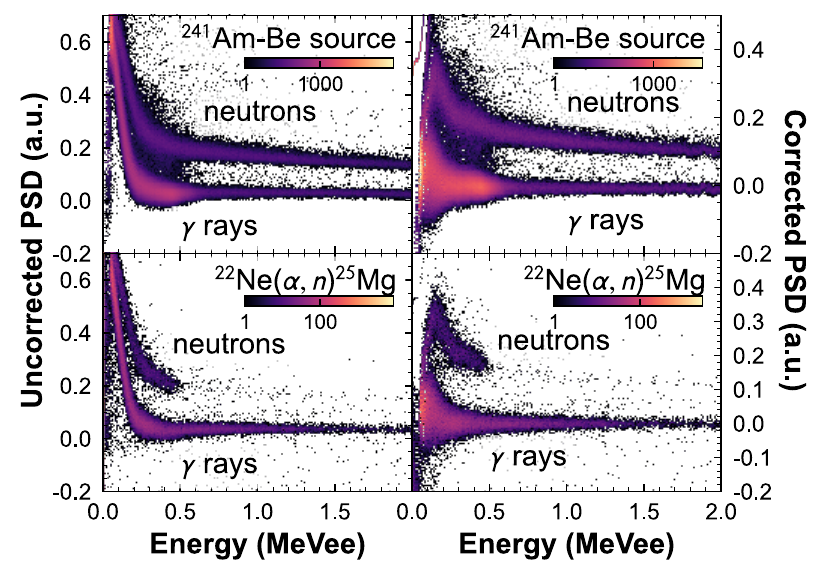}}
\caption{
	\emph{Left panels:} Uncorrected PSD vs.\ total energy spectrum for a single DEMAND detector. The top-left panel shows measurements using a \nuc{241}{Am}Be calibration source, and the bottom-left panel shows in-beam measurements of the \Nean{} reaction. \emph{Right panels:} Corrected PSD vs.\ total energy for the same detector and neutron sources as the left panels.
}
\label{fig:PSD_uncorr}
\end{figure*}

\begin{figure*}
	\centering
	\centerline{\includegraphics[width=190mm]{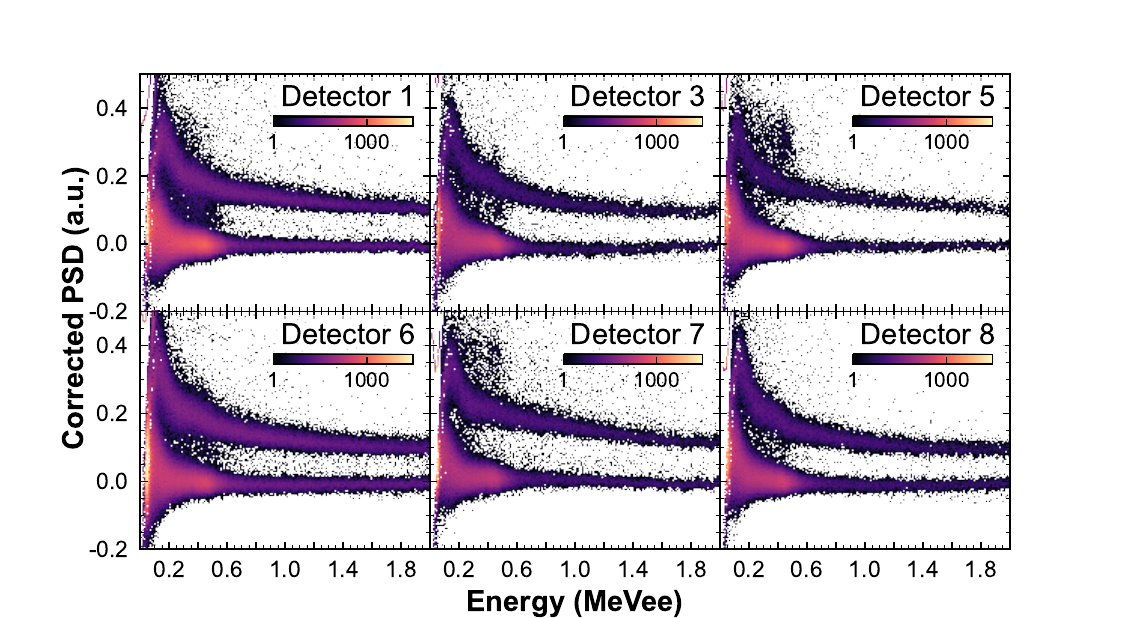}}
	\caption{
		Corrected PSD vs.\ total energy spectra for each of the six DEMAND detectors utilized in the experiment. This figure shows measurements using a \nuc{241}{Am}-Be calibration source.
	}
	\label{fig:PSD_corr_source}
\end{figure*}

\begin{figure*}
	\centering
	\centerline{\includegraphics[width=190mm]{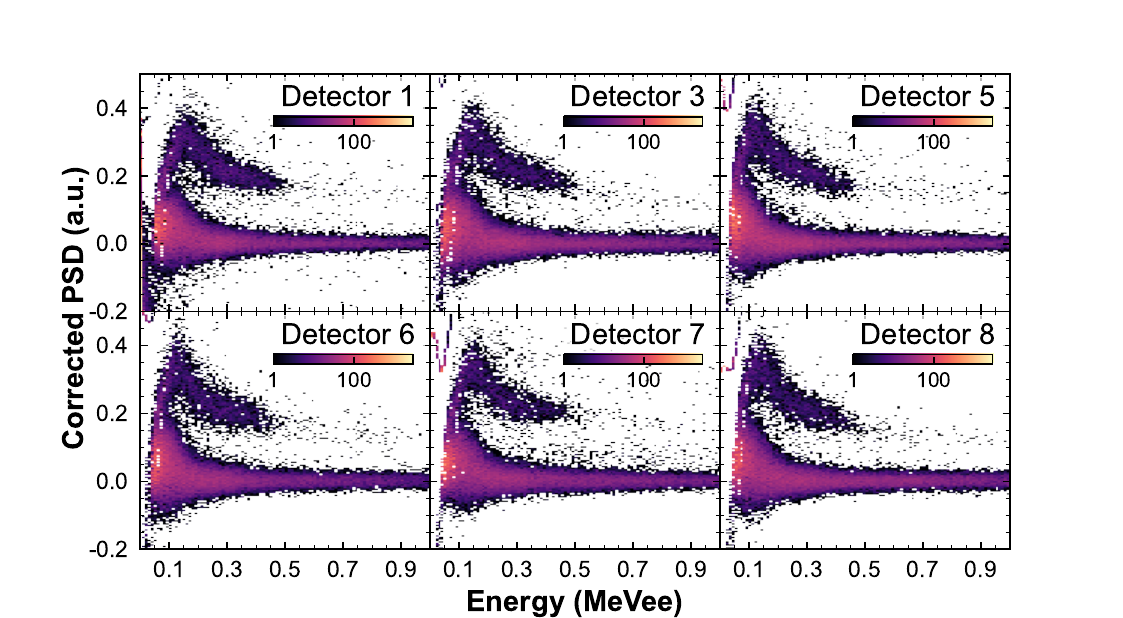}}
	\caption{
		Corrected PSD vs.\ total energy spectra for each of the six DEMAND detectors utilized in the experiment. This figure shows in-beam measurements of the \Nean{} reaction.
	}
	\label{fig:PSD_corr_beam}
\end{figure*}

A major advantage of using OGS over conventional plastic scintillators is its $n/\gamma$ discrimination capabilities using PSD. Many direct $(\alpha,n)$ or $(p,n)$ measurements \edit{will potentially} include a significant 
{{flux of beam-induced $\gamma$-rays incident on the neutron detectors, }}
either from strong radiative capture channels, or more seriously in the case of many radioactive-beam experiments, decay of beam particles implanted near the target. While time of flight and other recoil identification signals can suppress much of this $\gamma$-ray background, PSD offers a valuable \edit{additional means of background rejection} that is essential in high-background or low-yield measurements.

The PSD capabilities of the DEMAND detectors were evaluated using both in-beam \Nean{} data and ${}^{241}$AmBe source data. The PSD was determined from the recorded charge in the long and short gates (see Section~\ref{sec:expt}), with a PSD parameter constructed offline as
\[
	\mathrm{PSD} = 1 - E_{s}/E_{l},
\]
with $E_{s}$ and $E_{l}$ the charge recorded by the short and long gates, respectively. Since the short gate integrates only the first ${\sim} 10$-$15$ ns of the pulse (plus some leading DC offset), the PSD parameter effectively represents the ratio of charge in the pulse tail to the total charge. \editbf{This is referred to as the charge-integration method, and it is one of the most ubiquitous PSD techniques \cite{Knoll2010, Ranucci1995, Cester2014, Zhou2023}.}
Prior to constructing the PSD parameter, both $E_{l}$ and $E_{s}$ were calibrated into MeVee using the position of the \nuc{137}{Cs} Compton edge, as well as the position of the QDC pedestal (assumed to be zero energy). Figure~\ref{fig:PSD_uncorr} (left panels) displays the PSD parameter vs.\ $E_{l}$, for a single detector and both \Nean{} and \nuc{241}{Am}Be data. In both plots, a clear separation is seen between the top (neutron) and bottom ($\gamma$-ray) bands, with visible separation persisting down to ${\sim}100$~keVee. As expected, the \Nean{} neutrons are absent above $E_{l} \sim 0.5$~MeVee, the expected maximum energy deposition (electron-equivalent) for the ${\sim}1.65$~MeV neutrons that intersect the detectors.

In order to quantify the PSD quality of the detectors, the raw PSD spectra were corrected to remove the up-bend observed for small $E_{l}$ values. \edit{We attribute this up-bend to small inaccuracies in the calibration zero intercepts, which generate a $1/E_{l}$ shape in the PSD parameter. While the shape of the observed up-bend closely matches the expected $1/E_{l}$ dependence, we chose to perform the correction empirically. For the empirical correction,}  the $\gamma$-ray band was isolated using a contour gate, and then the mean PSD was estimated for a series of $E_{l}$ bins. This estimated curve of PSD vs.\ $E_{l}$, for $\gamma$-ray events only, was then fit with a cubic spline to determine a correction function, $f(E)$. The corrected PSD was then calculated as 
\[
	\mathrm{PSD}_c = \mathrm{PSD} - f(E_{l}).
\]
The right panels of Figure~\ref{fig:PSD_uncorr} show the corrected PSD vs.\ $E_l$, for the same detector and neutron sources as the left panels of the same figure.
Figures~\ref{fig:PSD_corr_source} and \ref{fig:PSD_corr_beam} show histograms of $\mathrm{PSD}_c$ vs.\ $E_{l}$, for all six detectors in the array, with Figure~\ref{fig:PSD_corr_source} displaying \nuc{241}{Am}Be data and Figure \ref{fig:PSD_corr_beam} displaying 
\Nean{} data. In the corrected spectra, the up-bend at low energies is removed, and the PSD is roughly uniform as a function of $E_{l}$.

\begin{figure*}
\centering
\centerline{\includegraphics[width=140mm]{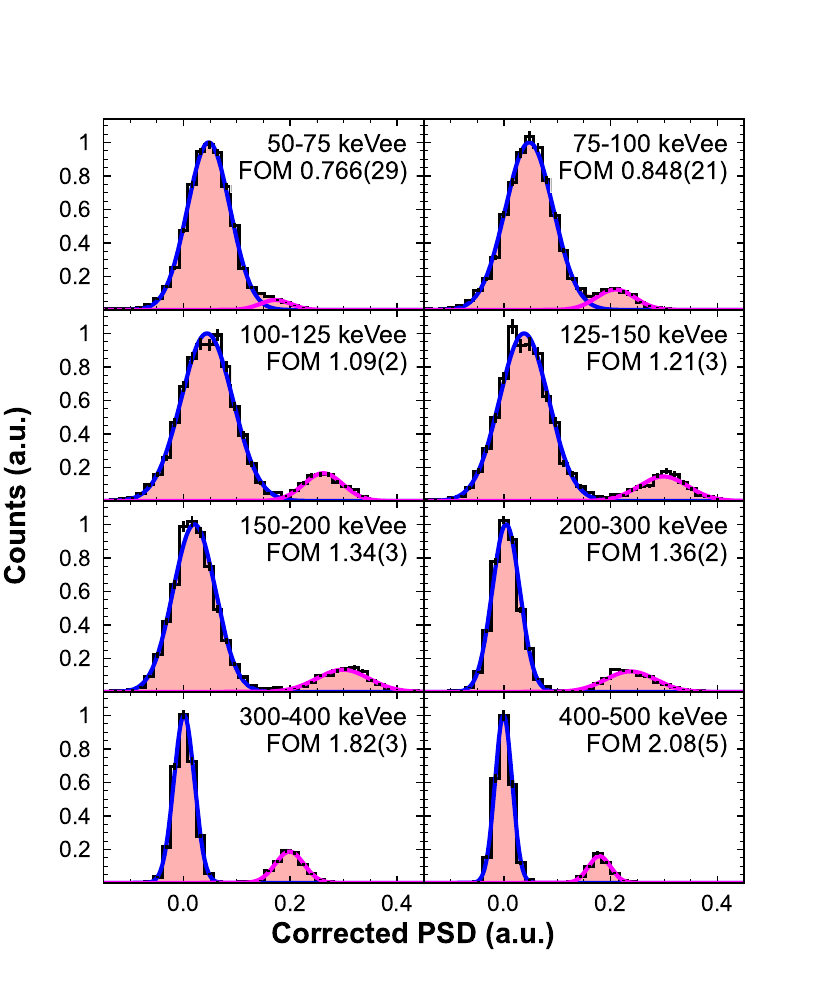}}
\caption{
	Sample fits of the PSD parameter using a \edit{two}-Gaussian function, for a variety of energy slices. The data and fits shown are from a single DEMAND detector \edit{(Detector 1)} and in-beam \Nean{} measurements. In each panel, the black histograms (with error bars) show the experimental data; red shaded curves show the total \edit{two}-Gaussian fit; blue and fuchsia solid curves show the individual Gaussian components. Each panel is labeled with its energy bin and the FOM resulting from the fit, including the uncertainty extracted from the fitting procedure. For display purposes, all data have been normalized such that the mode of the fit has a value of unity.
}
\label{fig:PSD_fits}
\end{figure*}

\begin{figure}
\centering
\centerline{\includegraphics[width=90mm]{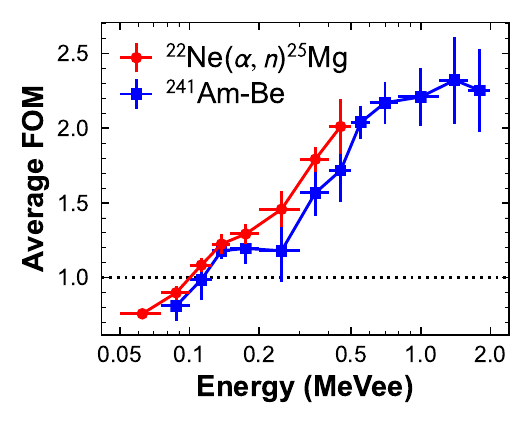}}
\caption{
	Average FOM vs.\ energy for all DEMAND detectors utilized in the experiment. The data points represent FOM values averaged across all six detectors, while the vertical error bars represent the sample standard deviation. Horizontal error bars denote the range of the energy bin for each projection. As indicated in the legend, the red filled circles represent in-beam \Nean{} reaction data, while blue filled circles represent \nuc{241}{Am}-Be source measurements.
}
\label{fig:FOM_vs_energy}
\end{figure}

To quantify the PSD quality, the two-dimensional spectra in Figures~\ref{fig:PSD_corr_source} and \ref{fig:PSD_corr_beam} were divided into energy slices and projected onto the $\mathrm{PSD}_c$ axis. The projected spectra show two distinct peaks, one from the neutrons and the other from the $\gamma$~rays. We fit each projection with a \edit{two}-Gaussian function, where the amplitudes, means, and standard deviations of each Gaussian were allowed to vary freely. The projected $\mathrm{PSD}_c$ spectra, along with the \edit{two}-Gaussian fit results across all energy bins are shown in Figure \ref{fig:PSD_fits}. The figure displays data from a single sample detector \edit{(Detector 1)} and in-beam \Nean{} radiation. As can be seen in the figure, the \edit{two}-Gaussian model is an excellent match to the data for all energy bins.

The results of the \edit{two}-Gaussian were used to determine a figure of merit (FOM), defined in terms of the individual means ($\mu_{n,\gamma}$) and standard deviations ($\sigma_{n,\gamma}$) of the neutron and $\gamma$-ray peaks:
\begin{equation}
	\mathrm{FOM} = \frac{\mu_n - \mu_\gamma}{
		2.355\cdot(\sigma_n+\sigma_\gamma)
	}.
\end{equation}
Defined as such, the FOM effectively gives the peak separation in units of the \edit{FWHM} \editbf{\cite{Cester2014, Annand1987, Perello2019}.} \edit{A common heuristic for good separation is  $\mathrm{FOM}>1$}, which corresponds to a $\lesssim 2\%$ overlap between the  peaks \editbf{\cite{Perello2019}}. To quantify the PSD quality as a function of energy deposition, we determined FOM values across all detectors and energy bins, and then plotted the average FOM vs.\ energy across the complete array. This was done both for \Nean{} data, where neutron energy deposition is limited to $\leq 500$~keVee, as well as \nuc{241}{Am}Be data, where neutrons deposit energy up to ${\sim}2$~MeVee. A plot of FOM vs.\ energy is shown in Figure~\ref{fig:FOM_vs_energy}, with \Nean{} and \nuc{241}{Am}Be data shown as separate curves. As expected, the FOM increases with deposited energy, with the $\mathrm{FOM}>1$ threshold being reached around 100~keVee. \edit{We note that even when $\mathrm{FOM}<1$, reliable neutron counts can be extracted from the area under the \edit{two}-Gaussian fit, as discussed in the following paragraph. However, gates placed on the PSD spectrum and applied to other parameters will suffer from noticeable $\gamma$-ray background.}

In addition to determining the FOM, the \edit{two}-Gaussian fits can also be used to extract the total number of neutron counts recorded by the detectors. This is done by taking the integral of the neutron peak, which gives a measure of the total number of neutron counts in an energy bin. Because the $\gamma$-ray \edit{background} is accounted for using this method, the neutron counts can be extracted even for energy bins where $\mathrm{FOM}<1$. From the present \Nean{} data, we extract a total of $3.547(21)\times 10^{4}$ neutron counts, where the statistical uncertainty is propagated from the individual fit-parameter uncertainties in each bin. We emphasize that the uncertainties extracted with this method are statistical only. Determining the true neutron yield requires folding the extracted counts with the efficiency curve, which introduces appreciable systematic uncertainties. \edit{The method demonstrated here to extract total neutron counts represents a first step towards extracting neutron-singles cross sections; however, a full cross section measurement is outside the scope of this article.}

\section{Monte Carlo Simulations}
\label{sec:MC}

\subsection{Overview}

To extract \edit{$(\alpha,n)$ or $(p,n)$} cross sections from measured yield data, the neutron detection efficiency must be well understood. \edit{For a given experiment, the neutron detection efficiency is a function of both the reaction kinematics and neutron scattering on passive beamline materials prior to entering the detectors. These effects are not possible to reproduce using sources, and hence Monte Carlo simulation is required to determine the detection efficiency for a reaction measurement.}
For organic scintillators, intrinsic detection efficiencies are dependent on the incident neutron energy, proton recoil quenching, \edit{light transport, and photocathode efficiency. The latter two effects can be incorporated empirically into the detector calibration, allowing simulated light-output spectra to be compared with measured data using the same effective detection threshold.} For inverse-kinematics measurements, the neutron detection efficiency is strongly dependent on the reaction kinematics, which impacts both the geometric efficiency (fraction of neutrons intersecting the detectors) and the incident neutron energy distribution. Neutron attenuation and scattering in passive components of the setup, e.g.\ beamline infrastructure, also plays a key role in determining the total detection efficiency.

In order to determine the efficiency of the DEMAND setup, we have developed a GEANT4 \cite{Agostinelli2003} Monte Carlo simulation of the DEMAND array.  The simulation integrates the DEMAND detector assembly into a new DRAGON GEANT4 simulation that is currently under development as a replacement for the existing GEANT3 simulation \cite{G3DRAGON}. 
The simulation includes the full beamline and detector geometry (beam pipes, BGO detectors, and gas target), while omitting other structures such as pumps, support hardware, and surrounding infrastructure. The origins and four-momenta of primary neutron tracks can either be generated from user-selected probability distributions or by importing saved event-by-event information from a DRAGON GEANT3 simulation.  The latter option allows for neutrons from arbitrary $(\alpha,n)$ or $(p,n)$ reactions to be generated while ensuring realistic beam dynamics (e.g., emittance and straggling in the gas target). It also allows for later correlation of the DEMAND simulation results with recoil information from GEANT3.

The primary neutrons were tracked down to thermal energies using the \texttt{FTFP\_BERT\_HP} physics list, which employs data-driven high-precision neutron interaction models below 20~MeV. The neutron HP package was configured to use evaluated data taken from the ENDF/B-VIII.0 library.
Neutron interactions within the OGS detectors generate energetic charged recoils, which were tracked through the active volume. The deposited energy in a given crystal was summed after applying a recoil-dependent quenching factor. For all recoil particles except protons \edit{(e.g., \nuc{12}{C} or light particles emitted from neutron-induced breakup reactions on carbon)}, we used the default GEANT4 Birks quenching factor, while for protons we instead used measured quenching curve reported by Laplace \emph{et al}.\ \cite{Laplace2020}. \edit{The Laplace \emph{et al.} data were taken with a $5.08~\mathrm{cm}\times 5.08~\mathrm{cm}$ cylindrical OGS scintillator, and while this is not identical to the present geometry, it represents the most similar measurement to date; in the next phase of the detector characterizations, we plan to augment the simulation with quenching curves measured using our cubical detectors.} The quenched energies were then given a random Gaussian spread to simulate the detector energy resolution, with the spread FWHM taken from the fits to source data discussed in Section~\ref{sec:EnergyRes}.  Events with post-quenching and post-smearing deposited energies falling above a user-selectable, detector-dependent threshold were classified as detected, while those falling below threshold were discarded. The detection efficiency is calculated as the ratio of detected to generated neutron events ($N_{det}/N_{gen}$), with the statistical uncertainty taken as $\sqrt{N_{det}}/N_{gen}$. Optionally, $N_{det}$ may be defined with a condition on the recoil species depositing the maximum energy into the crystal. This may be used, for example, to simulate the effect of a PID cut by selecting for proton recoils.
In addition to the efficiency calculation, the event-by-event origins and four-momenta of the primary neutrons, the deposited energies in each active crystal, and the recoil species with the maximum energy deposition were saved to disk for further analysis.

\subsection{Efficiency Curve}

\begin{figure}
\centering
\centerline{\includegraphics[width=90mm]{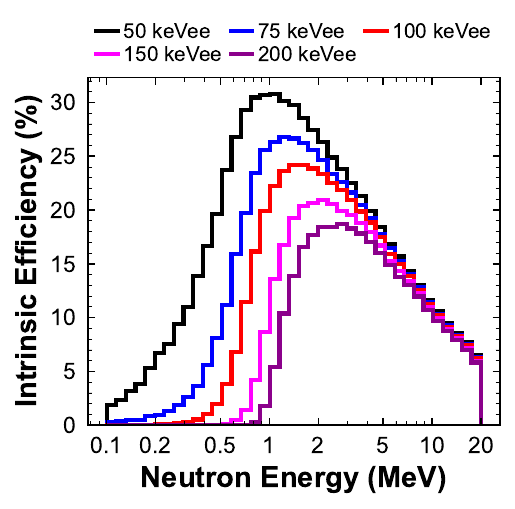}}
\caption{
	Simulated DEMAND efficiency curve as a function of neutron energy. The various histograms show the simulated efficiencies for different discriminator thresholds, as indicated in the legend.
}
\label{fig:effCurve}
\end{figure}

The simulation package can be used to generate intrinsic efficiency curves (as a function of incident neutron energy) for the DEMAND array.  This was done by generating primary neutrons with uniform random energies ranging from $0.1$--$20$~MeV and momentum vectors directly incident on the face of a single detector.\footnote{The $20$~MeV upper limit on the incident energies was chosen to match the upper limit of the GEANT4 high-precision neutron scattering libraries.} The detectors were housed in their standard thin-walled Al frame, but there was no other material between the primary neutrons and the detectors. The efficiency curve was calculated by binning the primary neutron energies logarithmically and plotting the ratio of detected to generated neutrons in each energy bin. A cut on proton recoils was included in the definition of detected neutrons, which results in a minor ($<3\%$) decrease in the efficiency for higher-energy incident neutrons.
Since the detection efficiency is strongly dependent on the energy threshold, simulations were performed with a variety of threshold settings: 50, 75, 100, 150, and 200~keVee.  The resulting efficiency curves at each threshold setting are shown in Figure~\ref{fig:effCurve}.

As expected, the simulated efficiency curves reflect a folding of the $n$--$p$ scattering cross sections (which decrease with increasing neutron energy) with the average energy transferred to the proton recoils. At low incident energies, the efficiency is primarily determined by the fraction of secondary protons above the detection threshold, leading the strongly threshold-dependent effects seen for this region in Figure~\ref{fig:effCurve}. For higher incident energies (above ${\sim}1$--$2$~MeV), the majority of secondary protons are above threshold, and the efficiency follows the shape of the scattering cross section. 
In a typical experiment, the detection threshold is expected to be in the range of $50$-$100$~keVee. For example, in the present experiment the $\mathrm{FOM}>1$ threshold was around $100$~keVee (see Section~\ref{sec:PSD}), and the estimated hardware threshold for contour-gated neutron events was $73$~keVee. For these threshold ranges, the efficiency peaks around $1$--$2$ MeV incident neutron energy, with maximum efficiency values ranging from ${\sim}20$--$30\%$. The minimum detectable neutron energy is also strongly threshold dependent, ranging from as low as ${\sim}0.2$~MeV for a $50$~keVee threshold to ${\sim}0.5$~MeV for a $100$~keVee threshold.

\subsection{\Nean{} Detection Efficiency}

\begin{figure}
\centering
\centerline{\includegraphics[width=90mm]{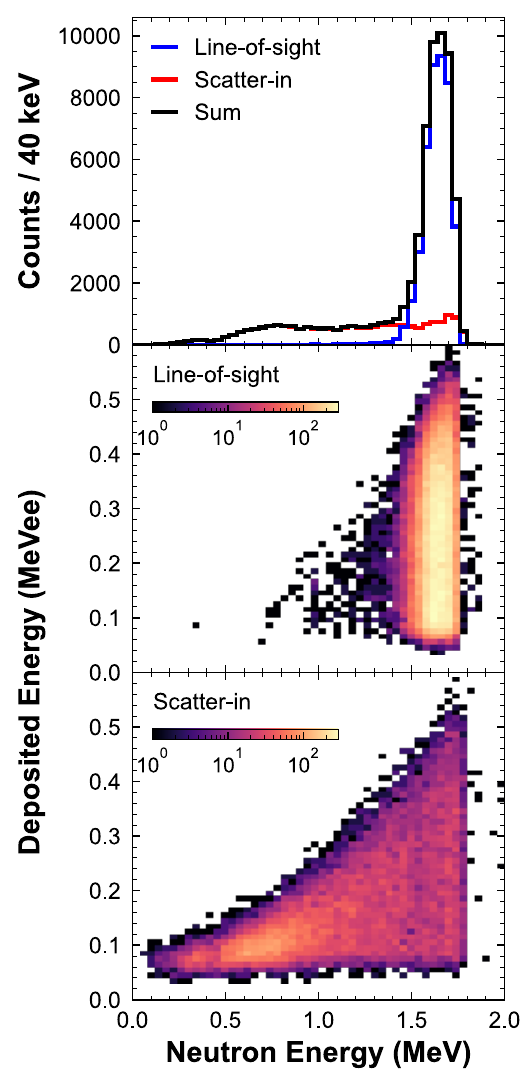}}
\caption{
	Summary of simulated neutron energy distributions for line-of-sight and scatter-in events (see text for the definition of line-of-sight vs.\ scatter-in).  The top panel shows the neutron kinetic energy distributions for both classes of event, as well as their sum.
The middle panel shows the deposited energy vs.\ neutron kinetic energy crossing the detector plane, for line-of-sight events, while the bottom panel shows the same for scatter-in events. 
}
\label{fig:G4energies}
\end{figure}

\begin{figure}
\centering
\centerline{\includegraphics[width=90mm]{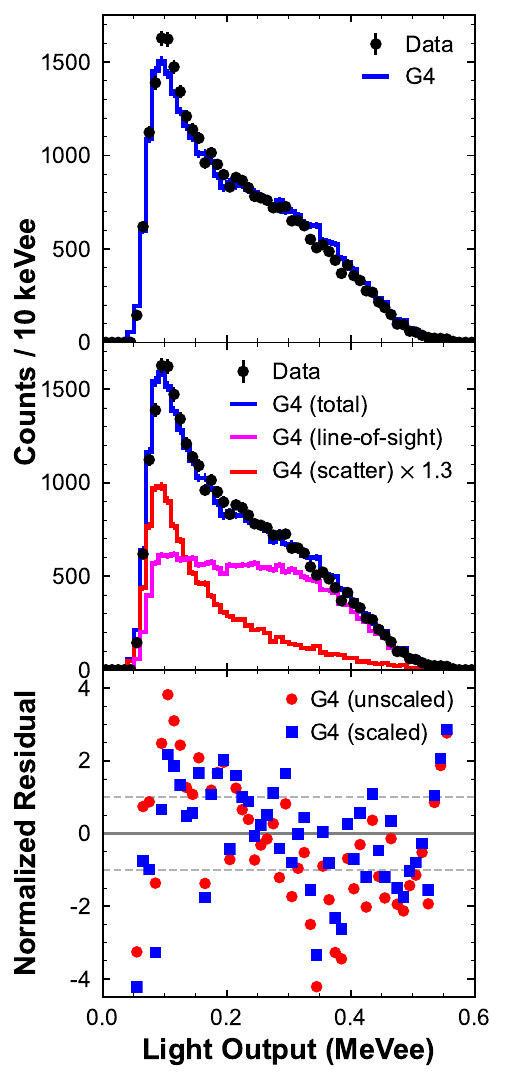}}
\caption{
	Comparison of GEANT4 simulations with measured energy-deposition spectra for neutron events. The top panel shows the measured data (black filled circles) together with the simulation results, scaled to the same total area. The \edit{middle} panel shows the data along with individual simulated contributions from line-of-sight neutrons (fuchsia) and neutrons that scatter into the detectors (red), scaled by a factor 1.3. The blue curve in the \edit{middle} panel shows the sum of these individual contributions. \edit{The bottom panel shows the normalized residual values for both the scaled and unscaled fits (see text).} 	\edit{The overall normalization of the simulated curves was a free parameter in the ML fit described in the text.}
}
\label{fig:G4}
\end{figure}

In addition to the efficiency curves, we also used the GEANT4 package to simulate the neutron-singles detection efficiency for the present \Nean{} experiment. For these simulations, the primary neutron tracks were taken from GEANT3 simulations of the \Nean{} reaction populating the $E_{cm} = 1213$~keV resonance with the same beam energy and target conditions observed during the experiment.  The simulated detector thresholds were set to 73~keVee, matching the threshold value extracted from a fit to the neutron-gated energy spectrum. Detectors 2 and 4 (as labeled in Table~\ref{tab:DetectorPositions}) were treated as inactive in the efficiency simulation. They were physically present and served as a scattering surface but were not treated as active detectors. This matches the experimental conditions during the initial set of runs whose analysis is reported in Section~\ref{sec:Analysis}. As with the efficiency curves, the definition of detected neutrons included a cut on proton recoils.

Due to the significant amount of beamline infrastructure surrounding the gas target, we anticipate that a substantial fraction of neutrons will be attenuated or scattered on passive materials. The impact of this passive scattering can be evaluated using the simulation, which includes both elastic and inelastic interactions between neutrons and the passive materials.
Analyzing the \Nean{} efficiency simulations, we find that a substantial fraction of the detected neutrons have initial trajectories that do not intersect with a DEMAND detector. Instead, they scatter into an active volume through some earlier interaction with passive material.
Furthermore, we find that the DEMAND energy-deposition spectrum provides a modest degree of sensitivity to the energies of the incident neutrons and hence their trajectories prior to entering the detector. This can be demonstrated by dividing simulated events into two categories: ``line-of-sight'' (initial trajectory intersects a detector) and ``scatter-in'' (initial trajectory does not intersect a detector).
For line-of-sight events, the neutron trajectories  are mostly undisturbed prior to entering the detector, and hence the incident neutron energies are concentrated near the initial reaction energy of ${\sim}1.65$~MeV. In contrast, scatter-in events are concentrated at lower incident energies since their kinetic energy is reduced by the passive scattering processes.

The top panel of Figure~\ref{fig:G4energies} shows the incident neutron kinetic energy distributions (as the neutron crosses the detector plane), for both line-of-sight and scatter-in events, as well as their sum.  As expected, the line-of-sight events are strongly peaked near the ${\sim}1.65$~MeV kinetic energy expected for \Nean{} neutrons intersecting a detector. In contrast, the scatter-in events display a broad incident energy distribution extending down to the detection threshold.  The bottom two panels of Figure~\ref{fig:G4energies} show the simulated post-quenching and post-smearing deposited energies vs.\ the incident neutron kinetic energies, for line-of-sight and scatter-in events as indicated. Here we see that the line-of-sight events give rise to a broad deposited-energy distribution, extending up to ${\sim}0.5$~MeVee, as expected for ${\sim}1.65$~MeV, quasi-monoenergetic incident neutrons.  In contrast, the broad and lower-energy incident energy distribution for scatter-in events gives rise to an energy-deposition spectrum that is more sharply peaked around $0.1$~MeVee.

By comparing the simulated energy-deposition spectra with those measured during the experiment, we can infer the proportion of total detected events that were scatter-in vs.\ line-of-sight.  The top panel of Figure~\ref{fig:G4} compares the total simulated and measured energy deposition spectra. While the simulation results are generally in good agreement with the data, they slightly under-predict the peak near 0.1~MeVee.  This under-prediction has an impact on the simulated detection efficiency, potentially biasing the simulation results towards smaller efficiencies than reality. 
We obtain improved agreement between the simulation and data by scaling the scatter-in component so that the simulated energy-deposition spectrum best matches the data. Such a \emph{post hoc} scaling is justified if the additional scattering arises from, \edit{e.g.,} beamline infrastructure omitted from the simulation. \edit{While the precise source of the missing scatter-in is speculative at present, adjusting the scatter-in component allows us to estimate its effect, while conservatively assigning a systematic uncertainty to the missing scatter-in components as described later in this section.}
A Poisson maximum-likelihood (ML) fit, \edit{which treats both the scatter-in scaling factor and the overall normalization as free parameters}, yields a best-fit scaling factor of 1.3(1). The resulting scaled simulation, along with the individual components, is compared with the measured data in the \edit{middle} panel of Fig.~\ref{fig:G4}. \edit{The bottom panel of the figure shows the normalized residuals for both the scaled and unscaled simulations, where the normalized residual gives the difference between the measured and simulated values in units of the standard deviation.}
Including the scaling of the scatter-in component, the simulated neutron-singles detection efficiency for the present setup is $\eta = 1.56(3)\%$, where the error bar includes both the Monte Carlo statistical uncertainty and the ML fit uncertainty on the scaling factor. In contrast, the simulated efficiency without any scaling is $1.42(1)\%$.  Due to the \emph{post hoc} nature of the scaling, we take the scaled efficiency of $1.56\%$ as the central value and fold the statistical uncertainty with a systematic uncertainty of $\Delta\eta = \pm 0.14\%$, the magnitude of the difference between the scaled and non-scaled efficiencies.
We have investigated other sources of systematic uncertainty on the efficiency by varying numerous simulation inputs that influence the positions and momenta of the primary neutrons. These include the incident \nuc{22}{Ne} beam properties (energy, position, and angle), target properties that influence the resonant reaction position (${}^{22}\mathrm{Ne}+\mathrm{He}$ stopping power, target pressure and temperature), and the detector threshold \edit{and energy calibration}. Each simulation input was varied by its estimated $\pm 0.5\sigma$ and $\pm1\sigma$ uncertainty, and the resulting curves of simulated efficiency vs.\ $n\sigma$ were fit with either a first- or second-order polynomial, depending on whether the efficiency variation was approximately monotonic or symmetric about the nominal value. The systematic uncertainty contribution from each parameter was then obtained by folding the fit results with the estimated parameter uncertainty distribution. The estimated systematic uncertainties resulting from this procedure are tabulated in Table~\ref{tab:syserr}, \edit{including the $1\sigma$ variations used for each parameter}. In addition, we fold into the error budget the $\pm0.03\%$ statistical uncertainty, the aforementioned $\pm 0.14\%$ uncertainty from the scatter-in scaling, and an estimated $10\%$ relative uncertainty related to the OGS density. Adding all uncertainties from Table~\ref{tab:syserr} in quadrature results in a final efficiency value of \edit{$\eta = 1.56(45)\%$}.  We emphasize that both the nominal efficiency and the uncertainty value are specific to the present experiment.  Any changes to the setup or the reaction kinematics will impact both the nominal efficiency and the estimated uncertainty. As a result, the Monte Carlo procedure described here will need to be repeated for  the analysis of future experiments.
\edit{We also emphasize that while the presently-simulated \Nean{} light-output spectrum agrees well with the data, the simulation has not been fully checked against incident-neutron data taken in a more controlled setting. In the future, we plan to take data at a neutron beam facility, which will allow for a more detailed validation of the simulation codes.}

\begin{table*}
\centering
\caption{Sources of absolute $(\Delta\eta)$  and relative $(\Delta\eta/\eta$) uncertainty in the Monte Carlo efficiency simulations. \edit{Where applicable, the $1\sigma$ variations listed in the second column represent the estimated $1\sigma$ uncertainties on each of the parameters that were varied in the GEANT4 simulations.} See text for \edit{additional} details.
}
\label{tab:syserr}
\begin{tabular}{cccc}
\hline
Source & \edit{$1\sigma$ Variation} & $\Delta\eta$ & $\Delta\eta/\eta$ \\
\hline
\hline

Statistical$^a$ & \edit{---} & $\pm0.03\%$ & $1.9\%$\\
Scatter-in scaling$^b$ & \edit{---} & $\pm0.14\%$ & $9.0\%$\\
OGS density & \edit{$\pm0.11$~g/cm$^3$} & $\pm0.16\%$ & $10\%$\\
Beam Energy  & \edit{$\pm0.012$~MeV} & $\pm0.13\%$ & $8.5\%$\\
Beam $x$ position & \edit{$\pm1$~mm} & $\pm0.11\%$ & $7.2\%$\\
Beam $y$ position & \edit{$\pm1$~mm} & $\pm0.18\%$ & $12\%$\\
Beam $x$-angle & \edit{$\pm6$~mrad} & $\pm0.19\%$ & $12\%$\\
Beam $y$-angle & \edit{$\pm6$~mrad} & $\pm0.07\%$ & $4.4\%$\\
Stopping Power & \edit{$10^{-14}$~eV~cm$^2$} & $\pm0.10\%$ & $6.6\%$\\
Target Pressure & \edit{$\pm0.1$~mbar} & $\pm0.01\%$ & $0.84\%$\\
Target Temperature & \edit{$\pm5$~K} & $\pm0.10\%$ & $6.2\%$\\
Detector Threshold & \edit{$\pm1$~keVee} & $\pm0.04\%$ & $2.9\%$\\
\edit{Energy Calibration}  & \edit{$\pm7.4\%$$^c$} & \edit{$\pm0.16\%$} & \edit{$10\%$}\\
\hline{Total} & \edit{---} & \edit{$\pm0.45\%$} & \edit{$29\%$}\\
\hline
\end{tabular}

\vspace{1mm}
\footnotesize{
$^a$Includes Monte Carlo statistics and uncertainty on the ML fit for the scatter-in scaling factor.
}\\
\footnotesize{
$^b$Magnitude of the difference between the scaled and unscaled efficiencies.
}\\
\footnotesize{
\edit{$^c$Estimated $1\sigma$ relative error on the energy calibration at the threshold value of $73.1$~keVee.}
}
\end{table*}

\section{Summary and Conclusions}

We have developed the new DEMAND array of OGS scintillators for use in inverse-kinematics $(\alpha,n)$ or $(p,n)$ experiments at DRAGON or other recoil separator facilities, commissioning the array with an in-beam measurement of the $E_{cm} = 1213$~keV \Nean{} resonance at DRAGON.
The DEMAND array is well-suited for deployment in compact geometries necessitated by recoil separator infrastructure. 
It is capable of detecting prompt neutrons with an estimated single-detector timing resolution of \
\edit{$640(30)$~ps FWHM at ${\sim}250$--600~keVee deposited energy}, making it well suited for identifying recoil--neutron coincidences using time of flight.
%
%
Using the OGS PSD, the array is capable of separating neutron and $\gamma$-ray events with $\mathrm{FOM}>1$ down to at least a $100$~keVee threshold. This provides a high degree of background $\gamma$-ray rejection while maintaining good detection efficiency for neutrons below 1 MeV.   Using GEANT4 simulations, we find that the intrinsic detection efficiencies can be as high as $30\%$, and depending on the detection threshold, the minimum detectable neutron energy ranges from ${\sim}0.2$--$0.5$~MeV. The GEANT4 simulations also allow us to determine neutron (or neutron-recoil) detection efficiencies for arbitrary reactions and detector geometries, including realistic treatment of passive neutron scattering off beamline infrastructure. The total uncertainties of these efficiency estimations are dominated by systematics, which for the present experiment are on the order of \edit{$\Delta\eta/\eta \sim 30\%$}. \edit{In the future, we plan to take data at a neutron-beam facility, which will allow for absolute-scale measurements of the neutron detection efficiency, as well as well-characterized incident-neutron data for benchmarking the simulation code.}

Following its commissioning, DEMAND (in the same configuration described here) has been used in a new measurement of the $E_{cm} = 704$~keV \Nean{} resonance at DRAGON, which is the main contributor to the $s$-process reaction rate. A subset of \edit{six} detectors has also been used during a measurement of the $\alpha$-process \rxn{93}{Rb}{\alpha}{n}{96}{Y} reaction, using the EMMA recoil separator \cite{Davids2005}.  Future plans for the array include measurements of the \rxn{38}{Ar}{\alpha}{n}{41}{Ca} reaction, which influences \nuc{41}{Ca} production in core-collapse supernovae \cite{Nishimura2026}, as well as \rxn{43}{K}{p}{n}{43}{Ca} reaction, a key determinant of astronomically-observable \nuc{43}{K} $\gamma$-ray lines from core-collapse supernovae \cite{Hermansen2020}.

\section*{Acknowledgments}
The authors are grateful to the staff at TRIUMF for their beam-delivery efforts throughout the experiment.
TRIUMF's core operations are supported via a contribution from the federal government through the National Research Council of Canada, and the Government of British Columbia provides building capital funds.
{DRAGON is generously funded through the Natural Sciences and Engineering Research Council of Canada (NSERC) award SAPPJ-2023-00039.}
Authors from Saint Mary's University acknowledge the support of NSERC awards SAPIN-2020-00052 and RGPAS-2020-00001. 
%
Authors from U.~Surrey were supported by the UK Science and Technology Facilities Council (STFC) via Grant Number ST/Y000358/1.
%
%
\section*{Declaration of generative AI and AI-assisted technologies in the manuscript preparation process}

During the preparation of this work, the authors used ChatGPT (OpenAI) as a writing and coding aid. The tool was used to assist with the phrasing of a small number of sentences and to help develop selected analysis, simulation, and plotting codes. After using this tool, the authors reviewed and edited its output and take full responsibility for the content of the published article.


\begin{thebibliography}{10}
\expandafter\ifx\csname url\endcsname\relax
  \def\url#1{\texttt{#1}}\fi
\expandafter\ifx\csname urlprefix\endcsname\relax\def\urlprefix{URL }\fi
\expandafter\ifx\csname href\endcsname\relax
  \def\href#1#2{#2} \def\path#1{#1}\fi

\bibitem{Bliss2018}
J.~Bliss, M.~Witt, A.~Arcones, F.~Montes, J.~Pereira, Survey of astrophysical
  conditions in neutrino-driven supernova ejecta nucleosynthesis, Astrophys.\
  J. 855 (2018) 135.
\newblock \href {https://doi.org/10.3847/1538-4357/aaadbe}
  {\path{doi:10.3847/1538-4357/aaadbe}}.

\bibitem{Frebel2018}
A.~{Frebel}, {From Nuclei to the Cosmos: Tracing Heavy-Element Production with
  the Oldest Stars}, Annu.\ Rev.\ Nucl.\ Part.\ Sci. 68 (2018) 237--269.
\newblock \href {https://doi.org/10.1146/annurev-nucl-101917-021141}
  {\path{doi:10.1146/annurev-nucl-101917-021141}}.

\bibitem{Bliss2020}
J.~Bliss, A.~Arcones, F.~Montes, J.~Pereira, Nuclear physics uncertainties in
  neutrino-driven, neutron-rich supernova ejecta, Phys.\ Rev.\ C 101 (2020)
  055807.
\newblock \href {https://doi.org/10.1103/PhysRevC.101.055807}
  {\path{doi:10.1103/PhysRevC.101.055807}}.

\bibitem{Lugaro2023}
M.~Lugaro, M.~Pignatari, R.~Reifarth, M.~Wiescher, The s process and beyond,
  Annu.\ Rev.\ Nucl.\ Part.\ Sci. 73 (2023) 315--340.
\newblock \href {https://doi.org/10.1146/annurev-nucl-102422-080857}
  {\path{doi:10.1146/annurev-nucl-102422-080857}}.

\bibitem{Hermansen2020}
K.~Hermansen, S.~M. Couch, L.~F. Roberts, H.~Schatz, M.~L. Warren, Reaction
  rate sensitivity of the production of $\gamma$-ray emitting isotopes in
  core-collapse supernovae, Astrophys.\ J. 901 (2020) 77.
\newblock \href {https://doi.org/10.3847/1538-4357/abafb5}
  {\path{doi:10.3847/1538-4357/abafb5}}.

\bibitem{Avila2017}
M.~Avila, K.~Rehm, S.~Almaraz-Calderon, A.~Ayangeakaa, C.~Dickerson,
  C.~Hoffman, C.~Jiang, B.~Kay, J.~Lai, O.~Nusair, R.~Pardo,
  D.~Santiago-Gonzalez, R.~Talwar, C.~Ugalde, Study of $(\alpha,p)$ and
  $(\alpha,n)$ reactions with a multi-sampling ionization chamber, Nucl.\
  Instrum.\ Methods Phys.\ Res., Sect.\ A 859 (2017) 63--68.
\newblock \href {https://doi.org/10.1016/j.nima.2017.03.060}
  {\path{doi:10.1016/j.nima.2017.03.060}}.

\bibitem{Jayatissa2023}
H.~Jayatissa, M.~L. Avila, K.~E. Rehm, P.~Mohr, Z.~Meisel, J.~Chen, C.~R.
  Hoffman, J.~Liang, C.~M\"uller-Gatermann, D.~Neto, W.~J. Ong, A.~Psaltis,
  D.~Santiago-Gonzalez, T.~L. Tang, C.~Ugalde, G.~Wilson, Study of the
  $^{22}\mathrm{Mg}$ waiting point relevant for x-ray burst nucleosynthesis via
  the $^{22}\mathrm{Mg}(\ensuremath{\alpha},p)^{25}\mathrm{Al}$ reaction,
  Phys.\ Rev.\ Lett. 131 (2023) 112701.
\newblock \href {https://doi.org/10.1103/PhysRevLett.131.112701}
  {\path{doi:10.1103/PhysRevLett.131.112701}}.

\bibitem{Blankstein2024}
D.~Blankstein, D.~W. Bardayan, P.~D. O'Malley, J.~Allen, T.~L. Bailey,
  C.~Boomershine, E.~Briley, D.~P. Burdette, L.~K. Callahan, S.~Carmichael,
  A.~M. Clark, A.~C. Dombos, O.~Olivas-Gomez, S.~L. Henderson, L.~Morales,
  A.~D. Nelson, J.~Wilkinson, First total cross section measurements of
  $^{25,26}\mathrm{Mg}(\ensuremath{\alpha},n)^{28,29}\mathrm{Si}$ via direct
  recoil detection, Phys.\ Rev.\ C 110 (2024) 045803.
\newblock \href {https://doi.org/10.1103/PhysRevC.110.045803}
  {\path{doi:10.1103/PhysRevC.110.045803}}.

\bibitem{Williams2025}
M.~Williams, C.~Angus, A.~M. Laird, B.~Davids, C.~A. Diget, A.~Fernandez, E.~J.
  Williams, A.~N. Andreyev, H.~Asch, A.~A. Avaa, G.~Bartram, S.~Chakraborty,
  I.~Dillmann, K.~Directo, D.~T. Doherty, E.~Geerlof, C.~J. Griffin, A.~Grimes,
  G.~Hackman, J.~Henderson, K.~Hudson, D.~Hufschmidt, J.~Jeong, M.~C.
  Jim\'enez~de Haro, V.~Karayonchev, A.~Katrusiak, A.~Lennarz, G.~Lotay,
  B.~Marlow, M.~S. Martin, S.~Moll\'o, F.~Montes, J.~R. Murias, J.~O'Neill,
  K.~Pak, C.~Paxman, L.~Pedro-Botet, A.~Psaltis, E.~Raleigh-Smith, D.~Rhodes,
  J.~S. Rojo, M.~Satrazani, T.~Sauvage, C.~Shenton, C.~E. Svensson, D.~Tam,
  L.~Wagner, D.~Yates, First measurement of a weak $r$-process reaction on a
  radioactive nucleus, Phys.\ Rev.\ Lett. 134 (2025) 112701.
\newblock \href {https://doi.org/10.1103/PhysRevLett.134.112701}
  {\path{doi:10.1103/PhysRevLett.134.112701}}.

\bibitem{Gastis2021}
P.~Gastis, G.~Perdikakis, G.~Berg, A.~Dombos, A.~Estrade, A.~Falduto, M.~Horoi,
  S.~Liddick, S.~Lipschutz, S.~Lyons, F.~Montes, A.~Palmisano, J.~Pereira,
  J.~Randhawa, T.~Redpath, M.~Redshaw, J.~Schmitt, J.~Sheehan, M.~Smith,
  P.~Tsintari, A.~Villari, K.~Wang, R.~Zegers, A technique for the study of
  $(p,n)$ reactions with unstable isotopes at energies relevant to
  astrophysics, Nucl.\ Instrum.\ Methods Phys.\ Res., Sect.\ A 985 (2021)
  164603.
\newblock \href {https://doi.org/10.1016/j.nima.2020.164603}
  {\path{doi:10.1016/j.nima.2020.164603}}.

\bibitem{Tsintari2025}
P.~Tsintari, N.~Dimitrakopoulos, R.~Garg, K.~Hermansen, C.~Marshall, F.~Montes,
  G.~Perdikakis, H.~Schatz, K.~Setoodehnia, H.~Arora, G.~P.~A. Berg,
  R.~Bhandari, J.~C. Blackmon, C.~R. Brune, K.~A. Chipps, M.~Couder, C.~Deibel,
  A.~Hood, M.~H. Gamage, R.~Jain, C.~Maher, S.~Miskovich, J.~Pereira,
  T.~Ruland, M.~S. Smith, M.~Smith, I.~Sultana, C.~Tinson, A.~Tsantiri,
  A.~Villari, L.~Wagner, R.~G.~T. Zegers, Machine learning enabled measurements
  of astrophysical ($p,n$) reactions with the {SECAR} recoil separator, Phys.
  Rev. Res. 7 (2025) 013074.
\newblock \href {https://doi.org/10.1103/PhysRevResearch.7.013074}
  {\path{doi:10.1103/PhysRevResearch.7.013074}}.

\bibitem{Carlson2016}
J.~S. Carlson, P.~L. Feng, Melt-cast organic glasses as high-efficiency fast
  neutron scintillators, Nucl.\ Instrum.\ Methods Phys.\ Res., Sect.\ A 832
  (2016) 152--157.
\newblock \href {https://doi.org/10.1016/j.nima.2016.06.116}
  {\path{doi:10.1016/j.nima.2016.06.116}}.

\bibitem{Carlson2017}
J.~S. Carlson, P.~Marleau, R.~A. Zarkesh, P.~L. Feng, Taking advantage of
  disorder: {S}mall-molecule organic glasses for radiation detection and
  particle discrimination, J. Am. Chem. Soc. 139~(28) (2017) 9621--9626.
\newblock \href {https://doi.org/10.1021/jacs.7b03989}
  {\path{doi:10.1021/jacs.7b03989}}.

\bibitem{Shin2019}
T.~H. Shin, P.~L. Feng, J.~S. Carlson, S.~D. Clarke, S.~A. Pozzi, Measured
  neutron light-output response for trans-stilbene and small-molecule organic
  glass scintillators, Nucl.\ Instrum.\ Methods Phys.\ Res., Sect.\ A 939
  (2019) 36--45.
\newblock \href {https://doi.org/10.1016/j.nima.2019.05.036}
  {\path{doi:10.1016/j.nima.2019.05.036}}.

\bibitem{Laplace2020}
T.~Laplace, B.~Goldblum, J.~Bevins, D.~Bleuel, E.~Bourret, J.~Brown,
  E.~Callaghan, J.~Carlson, P.~Feng, G.~Gabella, K.~Harrig, J.~Manfredi,
  C.~Moore, F.~Moretti, M.~Shinner, A.~Sweet, Z.~Sweger, Comparative
  scintillation performance of {EJ-309}, {EJ-276}, and a novel organic glass,
  JINST 15 (2020) P11020.
\newblock \href {https://doi.org/10.1088/1748-0221/15/11/P11020}
  {\path{doi:10.1088/1748-0221/15/11/P11020}}.

\bibitem{Ghia2021}
N.~P. Giha, W.~M. Steinberger, L.~Q. Nguyen, J.~S. Carlson, P.~L. Feng, S.~D.
  Clarke, S.~A. Pozzi, Organic glass scintillator bars with dual-ended readout,
  Nucl.\ Instrum.\ Methods Phys.\ Res., Sect.\ A 1014 (2021) 165676.
\newblock \href {https://doi.org/10.1016/j.nima.2021.165676}
  {\path{doi:10.1016/j.nima.2021.165676}}.

\bibitem{Warburton2021}
W.~K. Warburton, J.~S. Carlson, P.~L. Feng, Organic glass scintillator ({OGS})
  property comparisons to {S}tilbene, {EJ}-276 and {BC}-404, Nucl.\ Instrum.\
  Methods Phys.\ Res., Sect.\ A 1018 (2021) 165778.
\newblock \href {https://doi.org/10.1016/j.nima.2021.165778}
  {\path{doi:10.1016/j.nima.2021.165778}}.

\bibitem{Pozzi2026}
S.~Pozzi, E.~Schneider, E.~Neely, R.~Lopez, P.~Feng, S.~Clarke, Organic glass
  scintillators: {M}aterial properties, performance characterization, and
  applications, Radiation Measurements 193 (2026) 107627.
\newblock \href {https://doi.org/10.1016/j.radmeas.2026.107627}
  {\path{doi:10.1016/j.radmeas.2026.107627}}.

\bibitem{BlueshiftOptics}
{Blueshift Optics}, Organic glass scintillators for radiation detection,
  \url{https://blueshiftoptics.com/}, accessed: 2026-08-14.

\bibitem{Hutcheon2003}
D.~Hutcheon, S.~Bishop, L.~Buchmann, M.~Chatterjee, A.~Chen, J.~D'Auria,
  S.~Engel, D.~Gigliotti, U.~Greife, D.~Hunter, A.~Hussein, C.~Jewett, N.~Khan,
  M.~Lamey, A.~Laird, W.~Liu, A.~Olin, D.~Ottewell, J.~Rogers, G.~Roy,
  H.~Sprenger, C.~Wrede, The \{DRAGON\} facility for nuclear astrophysics at
  {TRIUMF-ISAC}: design, construction and operation, Nucl.\ Instrum.\ Methods
  Phys.\ Res., Sect.\ A 498~(1–3) (2003) 190 -- 210.
\newblock \href {https://doi.org/10.1016/S0168-9002(02)01990-3}
  {\path{doi:10.1016/S0168-9002(02)01990-3}}.

\bibitem{Laxdal2003400}
R.~Laxdal, Acceleration of radioactive ions, Nucl.\ Instrum.\ Methods Phys.\
  Res., Sect.\ B 204~(0) (2003) 400 -- 409.
\newblock \href {https://doi.org/10.1016/S0168-583X(02)02104-3}
  {\path{doi:10.1016/S0168-583X(02)02104-3}}.

\bibitem{jayamanna:02C711}
K.~Jayamanna, F.~Ames, G.~Cojocaru, R.~Baartman, P.~Bricault, R.~Dube,
  R.~Laxdal, M.~Marchetto, M.~MacDonald, P.~Schmor, G.~Wight, D.~Yuan, Off-line
  ion source terminal for {ISAC} at {TRIUMF}, Rev.\ Sci.\ Instrum. 79~(2)
  (2008) 02C711.
\newblock \href {https://doi.org/10.1063/1.2816928}
  {\path{doi:10.1063/1.2816928}}.

\bibitem{Christian2014}
G.~Christian, C.~Akers, D.~Connolly, J.~Fallis, D.~Hutcheon, K.~Olchanski,
  C.~Ruiz, Design and commissioning of a timestamp-based data acquisition
  system for the {DRAGON} recoil mass separator, Eur.\ Phys.\ J.\ A 50~(4)
  (2014).
\newblock \href {https://doi.org/10.1140/epja/i2014-14075-0}
  {\path{doi:10.1140/epja/i2014-14075-0}}.

\bibitem{Knoll2010}
G.~F. Knoll, Radiation Detection and Measurement, 4th Edition, John Wiley \&
  Sons, Hoboken, NJ, 2010.

\bibitem{Tilley2002}
D.~R. Tilley, C.~M. Cheves, J.~L. Godwin, G.~M. Hale, H.~M. Hofmann, J.~H.
  Kelley, C.~G. Sheu, H.~R. Weller, Energy levels of light nuclei {A} = 5, 6,
  7, Nucl.\ Phys.\ A 708 (2002) 3--163.
\newblock \href {https://doi.org/10.1016/S0375-9474(02)00597-3}
  {\path{doi:10.1016/S0375-9474(02)00597-3}}.

\bibitem{Basunia2015}
M.~S. Basunia, Nuclear data sheets for {A} = 22, Nucl.\ Data\ Sheets 127 (2015)
  69--190.
\newblock \href {https://doi.org/10.1016/j.nds.2015.07.002}
  {\path{doi:10.1016/j.nds.2015.07.002}}.

\bibitem{Browne2007}
E.~Browne, J.~K. Tuli, Nuclear data sheets for {A} = 137, Nucl.\ Data\ Sheets
  108 (2007) 2173--2318.
\newblock \href {https://doi.org/10.1016/j.nds.2007.09.002}
  {\path{doi:10.1016/j.nds.2007.09.002}}.

\bibitem{Kudomi1999}
N.~Kudomi, Energy calibration of plastic scintillators for low energy electrons
  by using {Compton} scatterings of gamma rays, Nucl.\ Instrum.\ Methods Phys.\
  Res., Sect.\ A 430 (1999) 96--99.
\newblock \href {https://doi.org/10.1016/S0168-9002(99)00200-4}
  {\path{doi:10.1016/S0168-9002(99)00200-4}}.

\bibitem{Stevanato2011}
L.~Stevanato, D.~Fabris, X.~Hao, M.~Lunardon, S.~Moretto, G.~Nebbia,
  S.~Pesente, L.~Sajo-Bohus, G.~Viesti, Light output of {EJ228} scintillation
  neutron detectors, Appl.\ Radiat.\ Isot. 69 (2011) 369--372.
\newblock \href {https://doi.org/10.1016/j.apradiso.2010.10.022}
  {\path{doi:10.1016/j.apradiso.2010.10.022}}.

\bibitem{Bishop2003}
S.~Bishop, R.~E. Azuma, L.~Buchmann, A.~A. Chen, M.~L. Chatterjee, J.~M.
  D'Auria, S.~Engel, D.~Gigliotti, U.~Greife, M.~Hernanz, D.~Hunter,
  A.~Hussein, D.~Hutcheon, C.~Jewett, J.~Jos{\'e}, J.~King, S.~Kubono, A.~M.
  Laird, M.~Lamey, R.~Lewis, W.~Liu, S.~Michimasa, A.~Olin, D.~Ottewell, P.~D.
  Parker, J.~G. Rogers, F.~Strieder, C.~Wrede, The
  $^{21}${Na}$(p,\gamma)^{22}${Mg} reaction and oxygen-neon novae, Phys.\ Rev.\
  Lett. 90 (2003) 162501.
\newblock \href {https://doi.org/10.1103/PhysRevLett.90.162501}
  {\path{doi:10.1103/PhysRevLett.90.162501}}.

\bibitem{Ruiz2006}
C.~Ruiz, A.~Parikh, J.~Jos{\'e}, L.~Buchmann, J.~A. Caggiano, A.~A. Chen, J.~A.
  Clark, H.~Crawford, B.~Davids, J.~M. D'Auria, C.~Davis, C.~Deibel,
  L.~Erikson, L.~Fogarty, D.~Frekers, U.~Greife, A.~Hussein, D.~A. Hutcheon,
  M.~Huyse, C.~Jewett, A.~M. Laird, R.~Lewis, P.~Mumby-Croft, A.~Olin, D.~F.
  Ottewell, C.~V. Ouellet, P.~Parker, J.~Pearson, G.~Ruprecht, M.~Trinczek,
  C.~Vockenhuber, C.~Wrede, Measurement of the $e_{\mathrm{c.m.}}=184$ kev
  resonance strength in the $^{26g}${Al}$(p,\gamma)^{27}${Si} reaction, Phys.\
  Rev.\ Lett. 96 (2006) 252501.
\newblock \href {https://doi.org/10.1103/PhysRevLett.96.252501}
  {\path{doi:10.1103/PhysRevLett.96.252501}}.

\bibitem{Lotay2016}
G.~Lotay, G.~Christian, C.~Ruiz, C.~Akers, D.~S. Burke, W.~N. Catford, A.~A.
  Chen, D.~Connolly, B.~Davids, J.~Fallis, U.~Hager, D.~A. Hutcheon, A.~Mahl,
  A.~Rojas, X.~Sun, Direct measurement of the astrophysical
  $^{38}${K}$(p,\gamma)^{39}${Ca} reaction and its influence on the production
  of nuclides toward the end point of nova nucleosynthesis, Phys.\ Rev.\ Lett.
  116 (2016) 132701.
\newblock \href {https://doi.org/10.1103/PhysRevLett.116.132701}
  {\path{doi:10.1103/PhysRevLett.116.132701}}.

\bibitem{Christian2025}
G.~Christian, D.~Hutcheon, I.~Casandjian, S.~Collins, A.~Edwin, E.~Desmarais,
  U.~Greife, A.~Katrusiak, A.~Lennarz, M.~Loria, S.~Moll\'o, J.~O'Connell,
  S.~Pascu, L.~Pedro-Botet, Z.~Podoly\'ak, B.~Reed, P.~Regan, C.~Ruiz,
  R.~Shearman, S.~Upadhyayula, L.~Wagner, M.~Williams, First in-beam
  demonstration of a hybrid {LaBr3/CeBr3/BGO} array to measure radiative
  capture resonance energies in an extended gas target using a novel time of
  flight technique, Nucl.\ Instrum.\ Methods Phys.\ Res., Sect.\ A 1072 (2025)
  170199.
\newblock \href {https://doi.org/10.1016/j.nima.2024.170199}
  {\path{doi:10.1016/j.nima.2024.170199}}.

\bibitem{Ranucci1995}
G.~Ranucci, An analytical approach to the evaluation of the pulse shape
  discrimination properties of scintillators, Nucl.\ Instrum.\ Methods Phys.\
  Res., Sect.\ A 354 (1995) 389--399.
\newblock \href {https://doi.org/10.1016/0168-9002(94)00886-8}
  {\path{doi:10.1016/0168-9002(94)00886-8}}.

\bibitem{Cester2014}
D.~Cester, M.~Lunardon, G.~Nebbia, L.~Stevanato, G.~Viesti, S.~Petrucci,
  C.~Tintori, Pulse shape discrimination with fast digitizers, Nucl.\ Instrum.\
  Methods Phys.\ Res., Sect.\ A 748 (2014) 33--38.
\newblock \href {https://doi.org/10.1016/j.nima.2014.02.032}
  {\path{doi:10.1016/j.nima.2014.02.032}}.

\bibitem{Zhou2023}
J.~Zhou, A.~Abdulaziz, Y.~Altmann, A.~Di~Fulvio, Generalized method for the
  optimization of pulse shape discrimination parameters, Nucl.\ Instrum.\
  Methods Phys.\ Res., Sect.\ A 1050 (2023) 168184.
\newblock \href {https://doi.org/10.1016/j.nima.2023.168184}
  {\path{doi:10.1016/j.nima.2023.168184}}.

\bibitem{Annand1987}
J.~R.~M. Annand, A fast module for pulse shape analysis, Nucl.\ Instrum.\
  Methods Phys.\ Res., Sect.\ A 262 (1987) 371--377.
\newblock \href {https://doi.org/10.1016/0168-9002(87)90876-X}
  {\path{doi:10.1016/0168-9002(87)90876-X}}.

\bibitem{Perello2019}
J.~F. Perello, S.~Almaraz-Calderon, B.~W. Asher, L.~T. Baby, N.~Gerken,
  K.~Hanselman, Characterization of the {CATRiNA} neutron detector system,
  Nucl.\ Instrum.\ Methods Phys.\ Res., Sect.\ A 930 (2019) 196--202.
\newblock \href {https://doi.org/10.1016/j.nima.2019.03.084}
  {\path{doi:10.1016/j.nima.2019.03.084}}.

\bibitem{Agostinelli2003}
S.~Agostinelli, J.~Allison, K.~Amako, J.~Apostolakis, H.~Araujo, P.~Arce,
  M.~Asai, D.~Axen, S.~Banerjee, G.~Barrand, F.~Behner, L.~Bellagamba,
  J.~Boudreau, L.~Broglia, A.~Brunengo, H.~Burkhardt, S.~Chauvie, J.~Chuma,
  R.~Chytracek, G.~Cooperman, G.~Cosmo, P.~Degtyarenko, A.~Dell'Acqua,
  G.~Depaola, D.~Dietrich, R.~Enami, A.~Feliciello, C.~Ferguson, H.~Fesefeldt,
  G.~Folger, F.~Foppiano, A.~Forti, S.~Garelli, S.~Giani, R.~Giannitrapani,
  D.~Gibin, J.~{G\'omez Cadenas}, I.~Gonz\'alez, G.~{Gracia Abril},
  G.~Greeniaus, W.~Greiner, V.~Grichine, A.~Grossheim, S.~Guatelli,
  P.~Gumplinger, R.~Hamatsu, K.~Hashimoto, H.~Hasui, A.~Heikkinen, A.~Howard,
  V.~Ivanchenko, A.~Johnson, F.~Jones, J.~Kallenbach, N.~Kanaya, M.~Kawabata,
  Y.~Kawabata, M.~Kawaguti, S.~Kelner, P.~Kent, A.~Kimura, T.~Kodama,
  R.~Kokoulin, M.~Kossov, H.~Kurashige, E.~Lamanna, T.~Lamp\'en, V.~Lara,
  V.~Lefebure, F.~Lei, M.~Liendl, W.~Lockman, F.~Longo, S.~Magni, M.~Maire,
  E.~Medernach, K.~Minamimoto, P.~{Mora de Freitas}, Y.~Morita, K.~Murakami,
  M.~Nagamatu, R.~Nartallo, P.~Nieminen, T.~Nishimura, K.~Ohtsubo, M.~Okamura,
  S.~O'Neale, Y.~Oohata, K.~Paech, J.~Perl, A.~Pfeiffer, M.~Pia, F.~Ranjard,
  A.~Rybin, S.~Sadilov, E.~{Di Salvo}, G.~Santin, T.~Sasaki, N.~Savvas,
  Y.~Sawada, S.~Scherer, S.~Sei, V.~Sirotenko, D.~Smith, N.~Starkov,
  H.~Stoecker, J.~Sulkimo, M.~Takahata, S.~Tanaka, E.~Tcherniaev, E.~{Safai
  Tehrani}, M.~Tropeano, P.~Truscott, H.~Uno, L.~Urban, P.~Urban, M.~Verderi,
  A.~Walkden, W.~Wander, H.~Weber, J.~Wellisch, T.~Wenaus, D.~Williams,
  D.~Wright, T.~Yamada, H.~Yoshida, D.~Zschiesche, Geant4—a simulation
  toolkit, Nucl.\ Instrum.\ Methods Phys.\ Res., Sect.\ A 506~(3) (2003)
  250--303.
\newblock \href {https://doi.org/10.1016/S0168-9002(03)01368-8}
  {\path{doi:10.1016/S0168-9002(03)01368-8}}.

\bibitem{G3DRAGON}
{DRAGON Collaboration}, {G3\_DRAGON},
  \url{https://github.com/DRAGON-Collaboration/G3_DRAGON}, commit 0b31fc8,
  accessed 2026-05-01 (2026).

\bibitem{Davids2005}
B.~Davids, C.~N. Davids, {EMMA:} {A} recoil mass spectrometer for {ISAC-II} at
  {TRIUMF}, Nucl.\ Instrum.\ Methods Phys.\ Res., Sect.\ A 544~(3) (2005)
  565--576.
\newblock \href {https://doi.org/10.1016/j.nima.2005.01.297}
  {\path{doi:10.1016/j.nima.2005.01.297}}.

\bibitem{Nishimura2026}
N.~Nishimura, C.~Fröhlich, T.~Rauscher, Uncertainties in the production of
  iron-group nuclides in core-collapse supernovae from {M}onte {C}arlo
  variations of reaction rates, Mon. Not. R. Astron. Soc. 546~(3) (2026)
  stag152.
\newblock \href {https://doi.org/10.1093/mnras/stag152}
  {\path{doi:10.1093/mnras/stag152}}.

\end{thebibliography}

\end{document}